\documentclass[]{aa} 

\usepackage{graphicx}
\usepackage{natbib}
\usepackage{longtable}
\usepackage{color}

\usepackage{txfonts}

\begin{document}

    \title{From precursor to~afterglow: \\The~complex evolution of~GRB\,210312B}
\author{    
M.~Jelínek\inst{1}\thanks{E-mail: mates@asu.cas.cz} \and
S.~A.~Grebenev\inst{2} \and
P.~Yu.~Minaev\inst{2} \and 
C.~C.~Th\"one\inst{1}\and
A.~de~Ugarte~Postigo\inst{3} \and
A.~Rossi\inst{4} \and
D.~Paris\inst{10} \and 
D.~A.~Kann \and 
J.~F.~Ag\"u\'i Fern\'andez\inst{7} \and
J.~Štrobl\inst{1} \and
A.~S.~Pozanenko\inst{2,8} \and
I.~V.~Chelovekov\inst{2} \and 
F.~Novotn\'y\inst{1} \and
S.~Karpov \inst{5} \and 
M.~Topinka\inst{6} \and
M.~Blažek\inst{7} \and
S.~V\'{\i}tek\inst{9} \and
R.~Hudec\inst{1,9}
}



\institute{
Astronomical Institute of the Czech Academy of Sciences (ASU-CAS), Fričova 298, 251 65 Ondřejov, Czechia \and
Space Research Institute of the Russian Academy of Sciences (IKI RAS), Profsoyuznaya 84/32, Moscow 117997, Russia \and
Artemis, Observatoire de la Côte d'Azur, CNRS, Nice 06300, France \and
INAF–Osservatorio di Astroﬁsica e Scienza dello Spazio, via Piero Gobetti 93/3, I-40129 Bologna, Italy \and
Institute of Physics of Czech Academy of Sciences, Prague, Czech Republic\and
School of Physics \& Astronomy, University of Birmingham, Birmingham B15 2TT, UK \and
Centro Astron\'omico Hispano en Andaluc\'ia, Observatorio de Calar Alto, Sierra de los Filabres, G\'ergal, Almer\'ia, 04550, Spain
\and
National Research University "Higher School of Economics", Myasnitskaya 20, Moscow 101000, Russia \and
Faculty of Electrical Engineering - FEE-CTU - ČVUT, 
Prague, Czech Republic \and
INAF–Osservatorio Astronomico di Roma, via Frascati 33, I-00040 Monte Porzio Catone, Italy
}


\date{Received March 31, 2021; accepted April 1, 2021}

\abstract { Long gamma-ray bursts (GRBs) are characterized by a brief gamma-ray
flash followed by a longer-lasting multiwavelength afterglow. The basic
mechanism is largely understood, and the early afterglow evolution often shows
complex features that provide crucial insights into the transition between
prompt and afterglow phases.} {We present a detailed analysis of GRB\,210312B,
detected by INTEGRAL, which exhibits both a precursor and a complex optical
afterglow evolution. Through careful modeling using Markov chain Monte Carlo
methods, we disentangled the contributions of an early optical flare and
forward shock emission.} {Our analysis reveals a gamma-ray precursor 17\,s
before the main pulse with a significantly softer spectrum (hardness ratio
$0.37 \pm 0.12$ versus $1.9 \pm 0.4$). The optical afterglow shows an early
peak at $76.0_{-5.1}^{+4.4}$ s characterized by a steep rise
($\alpha_\mathrm{flare,1} = -4.1_{-2.3}^{+1.6}$) and decay
($\alpha_\mathrm{flare,2} = 4.0_{-1.5}^{+2.1}$), followed by forward shock
emission with a broad hydrodynamic peak at around 150\,s.  In the subsequent
plateau phase, the afterglow initially has a complex structure before settling
into a final power law decay consistent with an electron distribution index $p
= 2.36_{-0.15}^{+0.18}$. The negligible host extinction ($A_\mathrm{V,host} =
-0.073_{-0.078}^{+0.100}$) suggests we are observing the intrinsic afterglow
spectrum. The host system consists of two luminous ($M_B \sim -21.7$)
components separated by 11.5 kpc at $z = 1.069$, which are possibly an
interacting galaxy pair.} {GRB\,210312B provides a rare opportunity to study
the prompt-to-afterglow transition in detail. The consistency of the forward
shock component with standard afterglow theory supports our physical
interpretation despite the lack of X-ray coverage.}


\maketitle

\section{Introduction}

Long gamma-ray bursts (GRBs) are the most luminous astrophysical phenomena
that occur at cosmological distances. With their spectra peaking in soft
gamma-rays, their isotropic luminosities can reach $10^{54}$\,erg\,s$^{-1}$,
enabling their detection across most of the observable Universe
\citep{woosley93, paczynski98}. Their origin has been traced to collapsing
massive stars that temporarily create conditions for a super-critically accreting
black hole, forming a relativistic jet-like outflow. These events are
characterized by an intense but brief flash of gamma rays followed by a
longer-lasting multiwavelength afterglow. While this basic picture is well
established, the transition between these phases and the early afterglow
evolution remain areas of active investigation, particularly as the ability to
capture these crucial early moments has improved dramatically with modern
facilities.

The short high-energy emission of GRBs is ascribed to internal
shocks within the jet \citep{Piran2004, Meszaros2006, Zhang2007,
KumarZhang2015}. External shocks, interactions between the ejected matter and
a preexisting circumburst material, are responsible for a longer-lasting
afterglow emission at lower frequencies. This afterglow emission produces a
synchrotron spectrum that can be described by several spectral power laws and
light curves that evolve as temporal power laws
\citep{MeszarosRees1997,Sari1998}.

The growing ability of very rapid follow-up by X-ray ({\it Swift}/XRT),
optical, and near-infrared instruments ({\it Swift}/UVOT and a large number of
ground-based telescopes) now permits us to observe the transition between
prompt emission and afterglow. In contrast to the simple afterglow behavior at
later times, in the early light curves we observe a range of features across
the electromagnetic spectrum: the onset of the afterglow, breaks to a steeper
or shallower decay, plateaus, and flares \citep{Zaninoni2013,
Swenson2013,Mazaeva2018IJMPD..2744012M}. The flares have mainly been studied in
X-rays, particularly with XRT on board the {\it
Swift} satellite, where a systematic follow-up of GRBs automatically starts
seconds after a gamma-ray detection. In X-rays, more than 30\% of all GRBs show
flaring activity \citep{Margutti2011, Yi2016}. Optical flare detections are
much rarer due to the paucity of very early observations. Correlations have
been found between rise and fall times, and later flares have been found to
have broader peaks in the optical \citep{Yi2017}. Many spectral and temporal
properties resemble those of spikes in the initial prompt $\gamma$-ray
emission, suggesting a similar origin for these later flares, such as a revived
central engine activity. Late flares have been suggested to be caused by a
sudden change in the interstellar medium; this had been ruled out by
theoretical models \citep{Gat2013}, but more recent modeling can explain late
flares with a reverse shock running into a stratified medium caused by previous
ejections from the star \citep{Ayache2020}.\ This helps explain late time
flaring, which is difficult to produce with a central engine activity.

In this paper we present multifrequency observations of GRB\,210312B, a
remarkable event and  one of the few bursts ever detected by the X-ray
telescope {INTErnational Gamma-Ray Astrophysics
Laboratory (INTEGRAL\/}) Joint European X-Ray Monitor (JEM-X). Despite having one of the faintest
afterglows ever studied in such detail, its optical light curve exhibits two
bright flares at very early times, providing a rare opportunity to study the
prompt-to-afterglow transition phase. 

In Sect.\,\ref{sect:obs} we present the
{INTEGRAL\/} high-energy observations and the comprehensive ground-based
follow-up of the afterglow. In Sect.\,\ref{sect:analysis} we detail our analysis
of the high-energy emission, characterize the optical afterglow evolution and
flares, and examine the afterglow spectroscopy and host galaxy properties.
Finally, in Sect.\,\ref{sect:conclusions} we discuss the implications of our
findings in the context of GRB physics. Throughout this paper, we use a
cosmological model with $H_0=67.3~\textnormal{km s}^{-1}\textnormal{Mpc}^{-1}$,
$\Omega_M=0.315$, and $\Omega_\Lambda=0.685$ \protect\citep{2014A&A...571A..16P}.

\section{Observations}\label{sect:obs}

\subsection{High-energy observations with INTEGRAL}

GRB\,210312B was discovered and localized \citep{mereghetti21}
with the automatic GRB triggering system the INTEGRAL Burst Alert System (IBAS; \citealt{mereghetti03,winkler03,kuulkers21} on March 12, 2021, at 20:52:17
UT (henceforth adopted as the trigger time, $T_0$, in our analysis). No detection
by other X-ray or gamma-ray experiments has been reported. 

The data from two INTEGRAL telescopes were used in our analysis of the
burst prompt X-ray and gamma-ray emission: IBIS and JEM-X. The two remaining instruments aboard
INTEGRAL --- the gamma-ray Spectrometer on
INTEGRAL (SPI) and  the spectrometer Anti Coincidence Shield (ACS) --- did
not provide any useful output. 

The IBIS  gamma-ray telescope of the observatory \citep{ubertini03} was designed
to map the sky and study detected sources in the hard X-ray/soft gamma-ray
energy ranges with an energy resolution $E/\Delta E\sim 13$ at 100 keV. The
imaging is based on the principle of a coded aperture. The telescope has a
$30^\circ\times30^{\circ}$ full width at zero response (FWZR) field of view
(FoV; the fully coded region is $9^{\circ}\times9^{\circ}$) and an angular
resolution of 12\arcmin\  (full width at half maximum). Within such a wide field
the telescope records $6-8$ bursts per year, while its resolution allows their
positions to be determined with an accuracy of $\leq2$\arcmin. The telescope has
two detector layers: the INTEGRAL Soft Gamma-Ray
Imager \citep[ISGRI;][]{lebrun03} with a maximum sensitivity in the range $18-200$ keV and the Pixellated Imaging Cesium Iodide Telescope
\citep[PICsIT;][]{labanti03}
sensitive in the range $0.2-10$ MeV. The area of each detector is $\sim 2600\
\mbox{cm}^2$, the effective area for events at the center of the FoV
is $\sim1100\ \mbox{cm}^2$ (half of the detectors are shadowed by the opaque
mask elements). The mentioned IBAS developed to quickly search for GRBs and timely report on them via the
Gamma-ray Coordinates Network (GCN) uses IBIS/ISGRI data for the
analysis. 

The JEM-X  X-ray telescope \citep{lund03} is sensitive in the 3-35 keV band.
This is also a coded-aperture telescope; its FoV with a diameter of
$13.\!\!^{\circ}2$ FWZR (the diameter of the fully coded region is
$4.\!\!^{\circ}8$) is bounded by a collimator. The detector is a gas chamber
with an entrance window area of $\sim490\ \mbox{cm}^2$ and an energy resolution
$\Delta E/E\sim 16$\% full width at half maximum (FWHM) at 6 keV. The effective area at the center of the FoV
is only $\sim75\ \mbox{cm}^2$, because more than 80\% of the detector is
shadowed by the opaque mask and collimator elements. There are two nearly
identical modules of the telescope aboard the observatory; if they operate
simultaneously, the effective area doubles to $\sim 150\ \mbox{cm}^2$. The
angular resolution of the JEM-X telescope ($\sim 3.\!\!^{\prime}35$ FWHM) is a
factor of 3 higher than that of the IBIS telescope.

\subsection{Optical observations}

\begin{figure} 
    \centering
    \includegraphics[width=0.707\hsize]{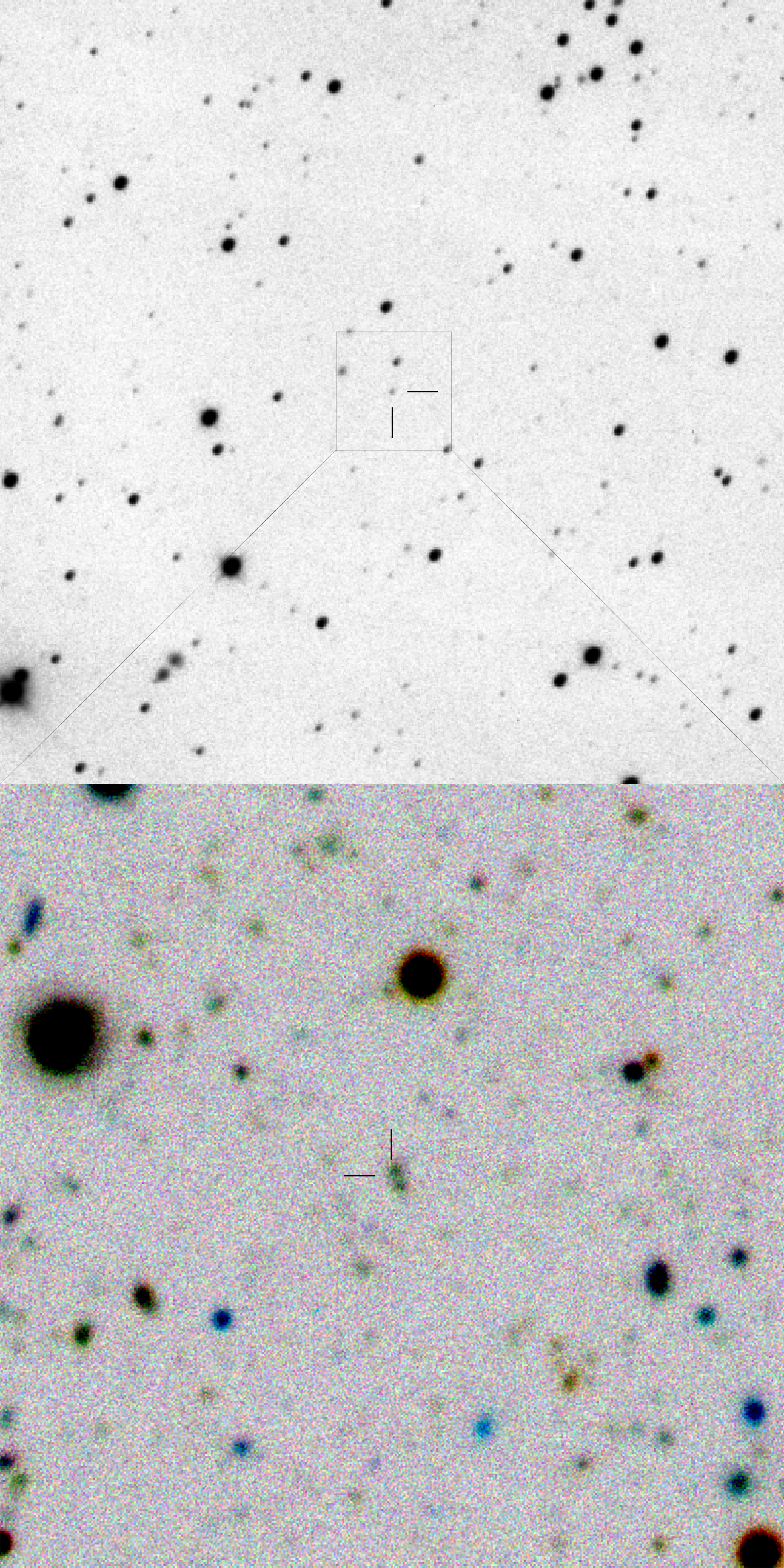}
    \caption{Top panel: Discovery image from the Ondřejov 0.5m telescope. The optical afterglow of GRB\,210312B is marked. The image covers {10\arcmin}
    per side. Bottom panel: $g r i$ image (value inverted) from the
    $2\times8.4$\,m LBT telescope. The location of the optical afterglow
    of GRB\,210312B is marked. The double structure of the underlying object is clearly
    visible. This image has a dimension of {$1.5\arcmin\times1.5$\,\arcmin}.
    \label{fig:opt_image}}
\end{figure}

The position of the event was promptly imaged by the D50 telescope in Ondřejov,
which reacted robotically to the trigger. An optical counterpart was identified
at \begin{center}$\alpha$=10:23:15.662$\pm$0.024
\quad$\delta$=+76:52:06.382$\pm$0.072 (J2000)\end{center} (155.815258
+76.868439; see our Fig.\,\ref{fig:opt_image} and \citealt{2021GCN.29651....1J}) and
later confirmed by \citet{2021GCN.29652....1J}. The D50 continued monitoring in
a clear filter calibrated to the $r$ band for about 0.5\,h after the GRB, until the
combination of fading afterglow and approaching cirrus clouds made further
observations impossible.

The analysis of the unfiltered D50 data is challenging in terms of calibration. 
First, the zero point needs to be related to a standard system, such as
Pan-STARRS (Panoramic Survey Telescope and Rapid Response System) AB magnitudes. This can be done by fitting the brightness of the
detected objects as a function of catalog stars. We performed a fit of
our unfiltered data as a function of $g$, $r$ and $i$ entries of the
ATLAS-Refcat2 \citep[Asteroid Terrestrial-impact Last Alert System, ][]{tonry18}. The D50 unfiltered photometry is best
fitted as $cr_\mathrm{raw} = r + 0.349 (g-r) - 0.839 (r-i)$. The brightness of
the object, however, corresponds to a new photometric system defined by the
width of its passband. To compare the afterglow brightness of this system
and that of standard filters, we needed to homogenize the result based on the known or
assumed color of the optical afterglow. In this case, we could calculate the
color indices $(g-r) = 0.373$ mag and $(r-i) = 0.256$ mag from the later
photometry; assuming a constant spectrum, we derived the relative shift of
the afterglow brightness ($r_\mathrm{OA} - cr_\mathrm{raw,OA} = +0.085$\,mag).
The unfiltered measurements provided in Appendix\,\ref{tab:phot} as a filter $cr$
include this correction and can be directly compared to the $r$ band, while
keeping in mind the assumption of a constant spectrum. 

While D50 was performing these early observations, the 2 m Perek telescope
joined the monitoring campaign approximately 0.5\,h after the trigger. Despite
increasingly challenging weather conditions at the Ondřejov observatory, it was
able to continue monitoring through thin clouds for about an hour longer than
D50, until denser clouds finally forced the termination of observations $\sim
1.5$\,h after the GRB.

We continued imaging observations for several days with the 2\,m Perek telescope
in Ondřejov in $g r i$ and $z$ filters, the 1.5\,m and 0.9\,m telescopes of the
Observatorio Sierra Nevada (OSN), Granada, Spain \citep{Kann2021GCN29716} in
BVRI filters, the 2.2\,m telescope at Calar Alto, Almeria, Spain, equipped with CAFOS in
BVRI and the 10.4\,m Gran Telescopio Canarias (GTC), equipped with OSIRIS, at
Observatorio Roque de los Muchachos, La Palma, Canary Islands, Spain
\citep{Kann2021GCN29653} in $r$. Finally, we obtained host-galaxy observations
at 56\,d after the GRB with the twin $2\times8.4$m Large Binocular Telescope
(LBT), Mt. Graham, Arizona, USA in Sloan $g' r' i' z'$. The LBT imaging data
were reduced using the data reduction pipeline developed at INAF - Osservatorio
Astronomico di Roma \citep{Fontana2014a}, and the photometry obtained is listed
in Appendix\,\ref{tab:phot} and Table\,\ref{tab:phot1}. 

\subsection{Optical spectroscopy}

The afterglow of GRB\,210312B was observed with OSIRIS \citep{Cepa2000} at the
GTC, at the Roque de los Muchachos Observatory (La Palma,
Spain) using the R1001B grism and a slit of $1^{\arcsec}$ aligned at
parallactic angle. This results in a spectral coverage between 3700 and 7800
\AA\/ with a resolving power of $\sim 610$. The spectroscopic observations were
performed in two runs. The first spectral exposure started on March, 12 2021, at
23:14:06 UT but was performed under adverse conditions, including bad seeing and
low transparency due to Calima. We obtained $2\times900$ s spectroscopic
exposures with the R1000B grism, after which the telescope had to be shut down
due to worsening conditions and only the first exposure was usable. The second
epoch started on at 01:50:25 UT and included $4\times900$ s spectra of somewhat
better quality.

Here we analyze the combined spectrum, which is an average of the first exposure
and the last four.
This $5\times900$ s spectrum has a mean epoch of 01:45:08
UT. A preliminary analysis of this spectrum was presented by
\citep{2021GCN.29655....1}, revealing a redshift of $z =1.069$.

\section{Analysis of the prompt and afterglow emission}\label{sect:analysis}

The analysis of the JEM-X and IBIS/PICsIT data is carried out with the standard
INTEGRAL software (OSA; version 10.2, \citealt{osa}), whereas the
analysis of the IBIS\-/ISGRI  data is performed with software developed at IKI
RAS (its basic principles aimed at carefully removing the nonuniform background
of the instrument and reconstructing the sky image were described by
\citealt{revnivtsev04}).

\subsection{GRB detection and localization}

\begin{figure} 
    \centering
    \includegraphics[width=0.9\hsize]{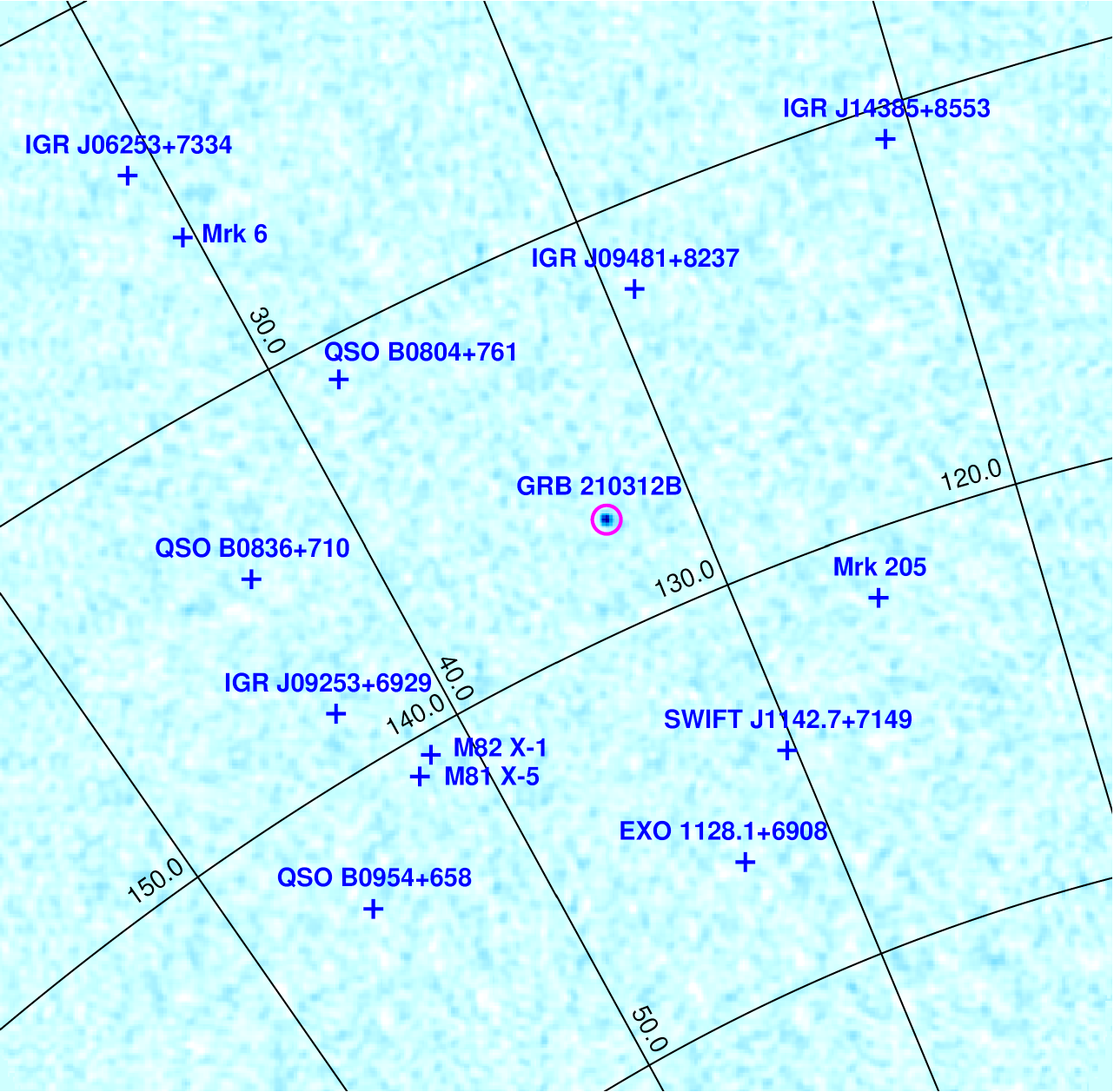}
    \caption{Sky image (S/N map) within the IBIS/ISGRI FoV
    obtained during the primary pulse of GRB\,210312B (2.4 s). Coordinates are
    Galactic. GRB\,210312B was the only source detected in the field (at a S/N
    level of $19.1\sigma$). The crosses indicate positions of known persistent
    X-ray sources (not detected in this short exposure).
    \label{fig:img_full}}
\end{figure}

Figure~\ref{fig:img_full} shows the sky image (S/N map) obtained by
IBIS/ISGRI in the 20-100 keV range during the $2.4$ s duration of the
GRB\,210312B primary pulse. There is only one (previously unknown) source
confidently (S/N $\simeq 19.1\sigma$) detected in the FoV, which we naturally
associate with GRB\,210312B. It is located at RA$\simeq 155.\!\!^{\circ}837$,
Dec. $\simeq76.\!\!^{\circ}874$ (epoch 2000.0, uncertainty $\sim45^{\arcsec}$).

\begin{figure}
    \centering
    \begin{minipage}{0.5\columnwidth}
	\includegraphics[width=1.0\columnwidth]{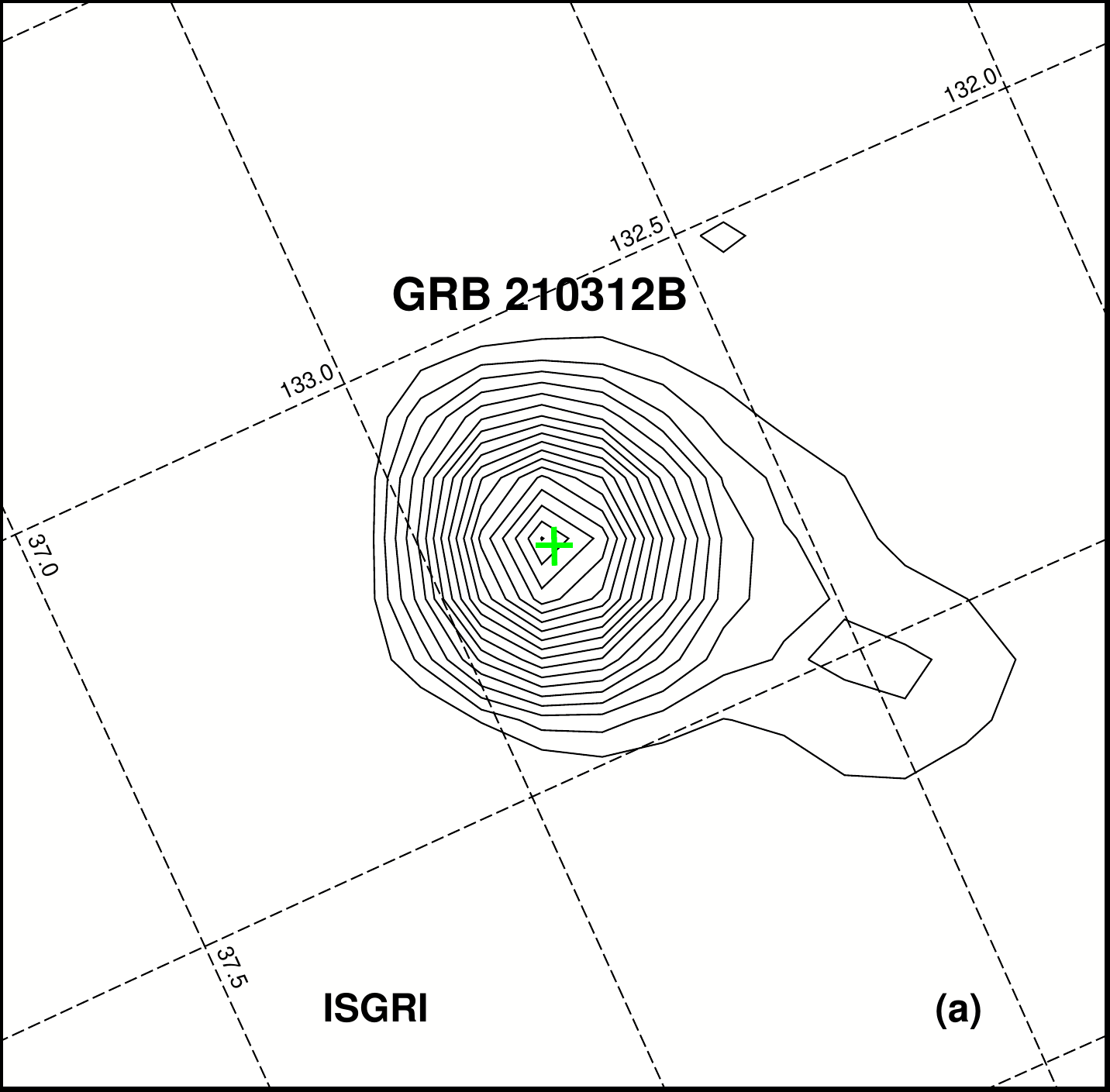}
    \end{minipage}\begin{minipage}{0.5\columnwidth}
	\includegraphics[width=1.\columnwidth]{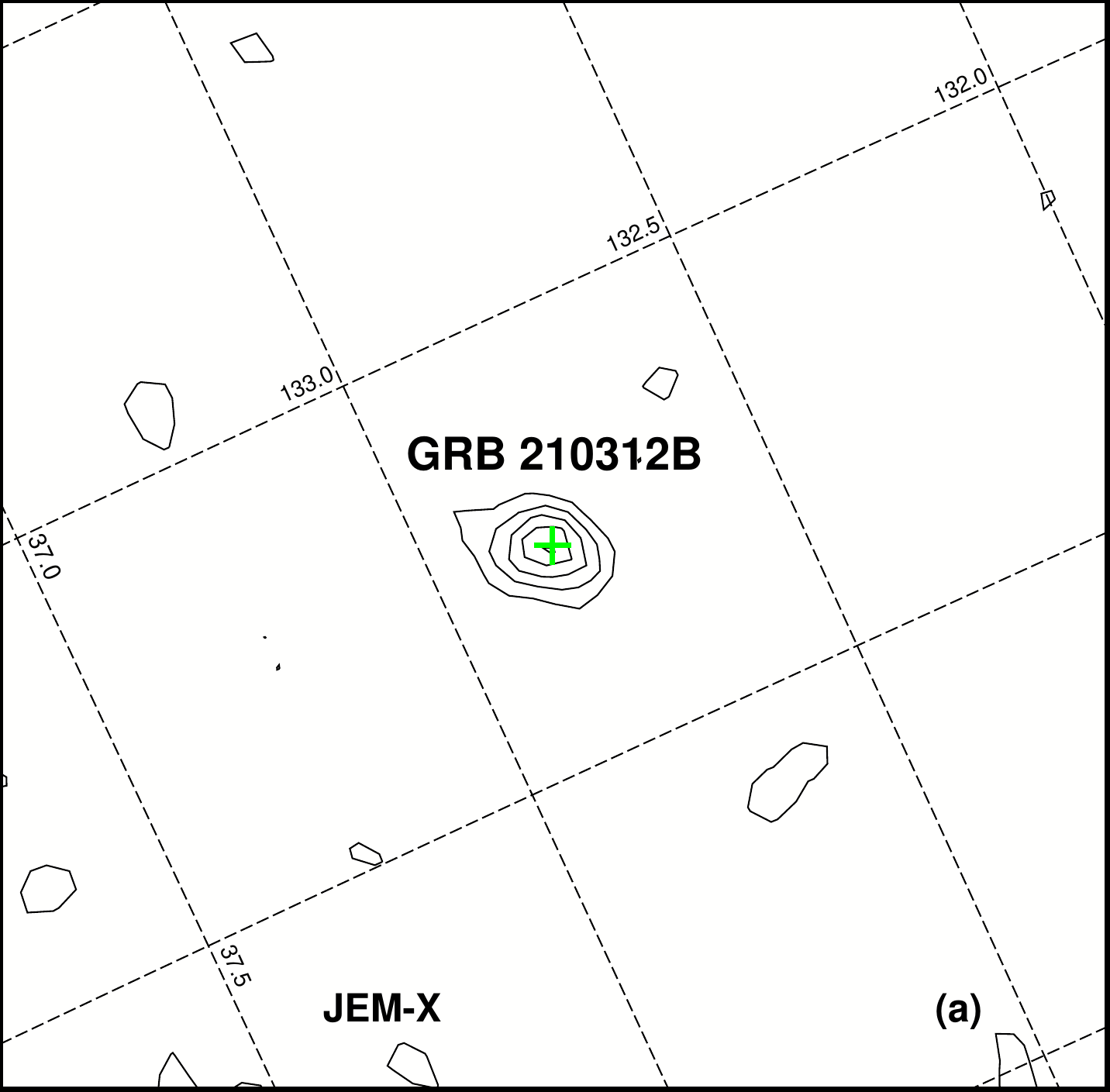}
    \end{minipage}
    \begin{minipage}{0.5\columnwidth}
	\includegraphics[width=1.0\columnwidth]{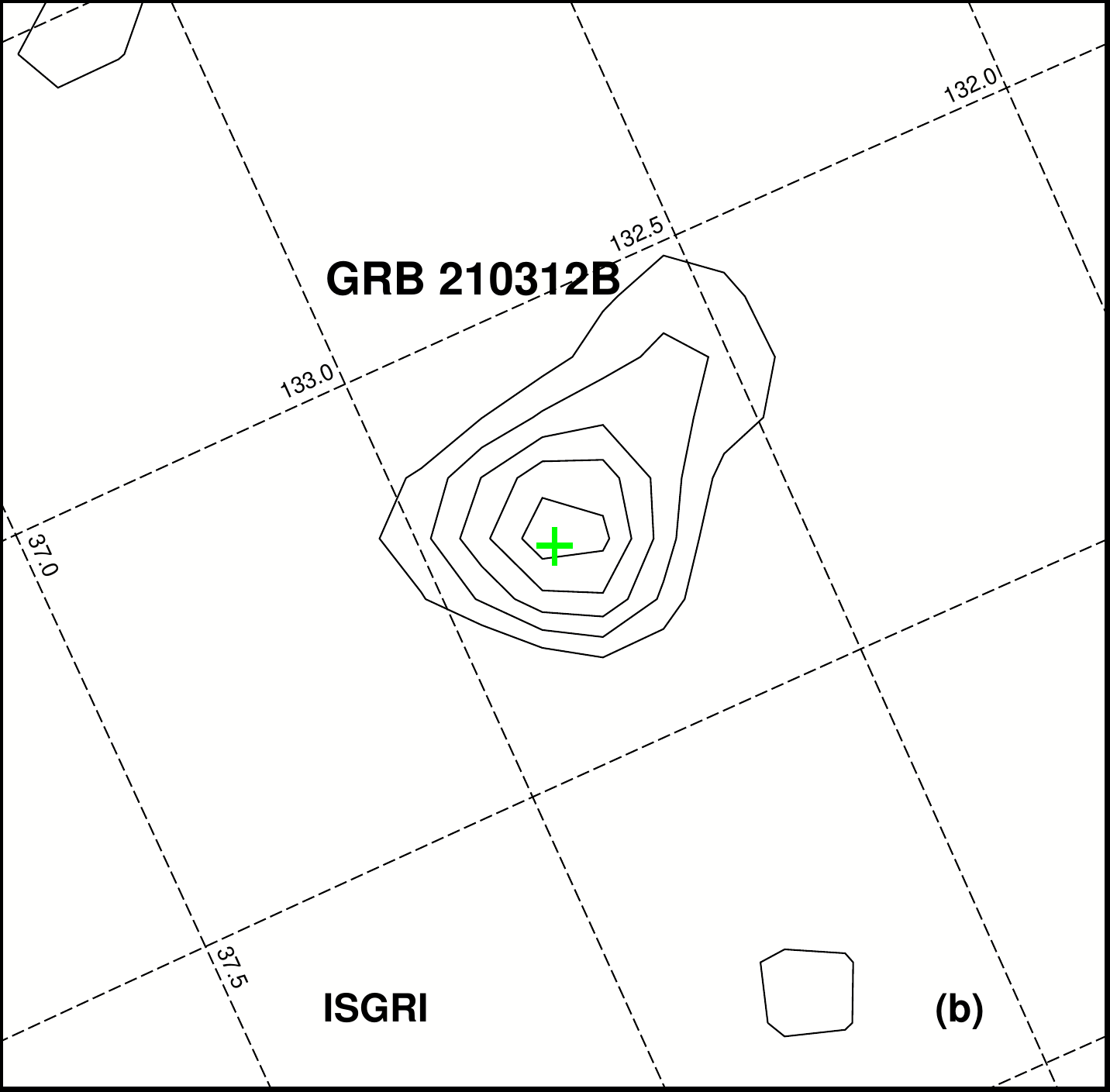}
    \end{minipage}\begin{minipage}{0.5\columnwidth}
	\includegraphics[width=1.0\columnwidth]{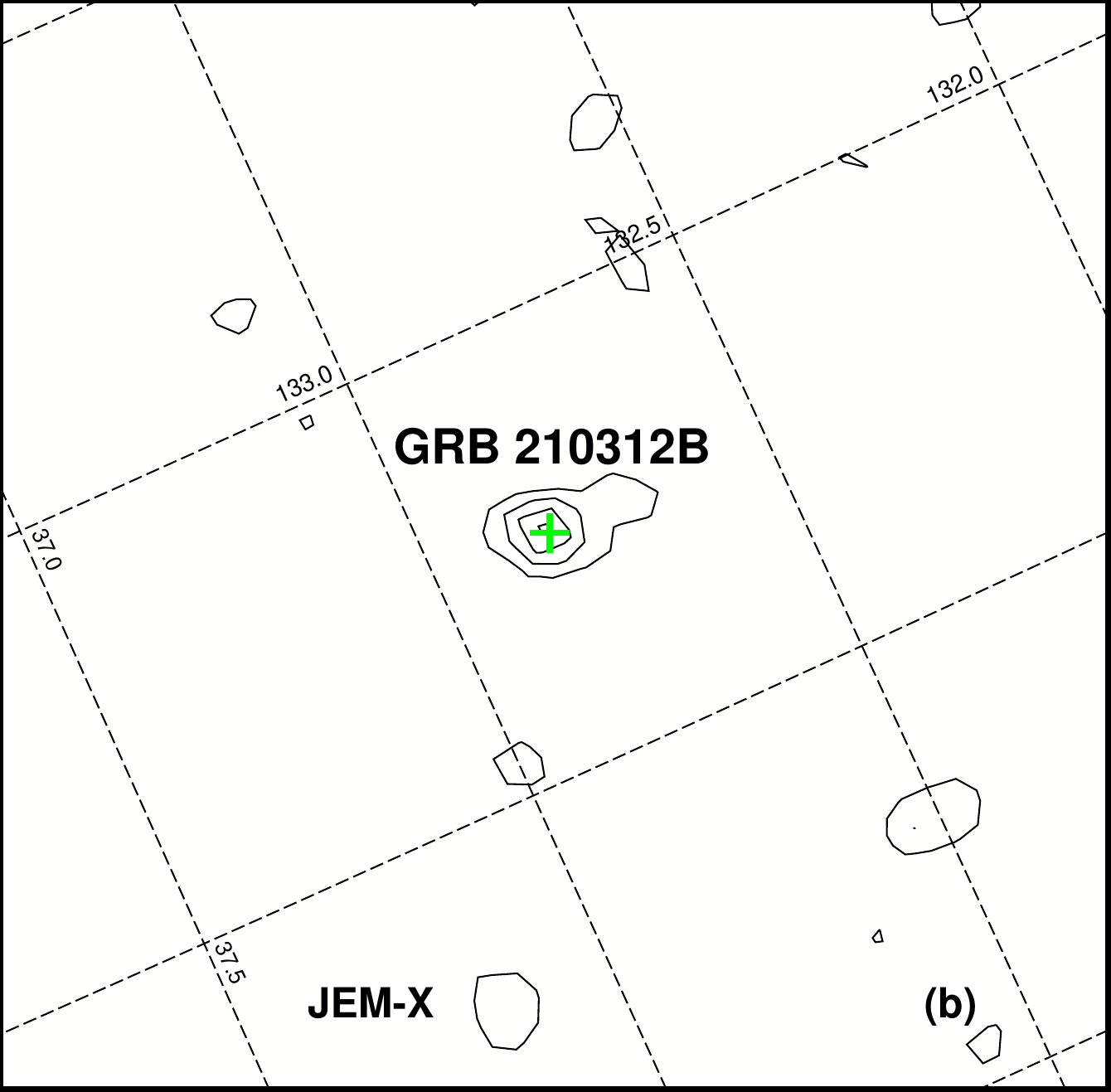}
    \end{minipage}
    \caption{Detection of the burst by INTEGRAL during the primary
    pulse (a) and precursor (b) phases. Shown are S/N maps of the sky region
    $1\fdg5\times 1\fdg5$ in size centered on the GRB\,210312B position. They
    were obtained by IBIS/ISGRI in the 20--100 keV range ({\sl left\/}; solid
    lines show the S/N levels at 2, 3, 4, ... $\sigma$) and JEM-X in the 3--20
    keV range ({\sl right\/}; the S/N levels are at 1, 2, 3, ... $\sigma$). The
    dashed lines show a grid of Galactic coordinates. The burst was detected
    with IBIS/ISGRI at $19.1\sigma$ during the primary peak and $6.6 \sigma$
    during the precursor, and with JEM-X at $5.2$ and $4.3 \sigma$,
    correspondingly. \label{fig:imgcr} }
\end{figure}

The source is also detected in the JEM-X $3-20$\,keV image (combining the data
from the two telescope units; see Fig.~\ref{fig:imgcr}a). Here, the maximum S/N  $\simeq 5.2\sigma$ was reached over a longer time interval of $\simeq 7.0$
s. 

\begin{figure} 
    \includegraphics[width=\columnwidth]{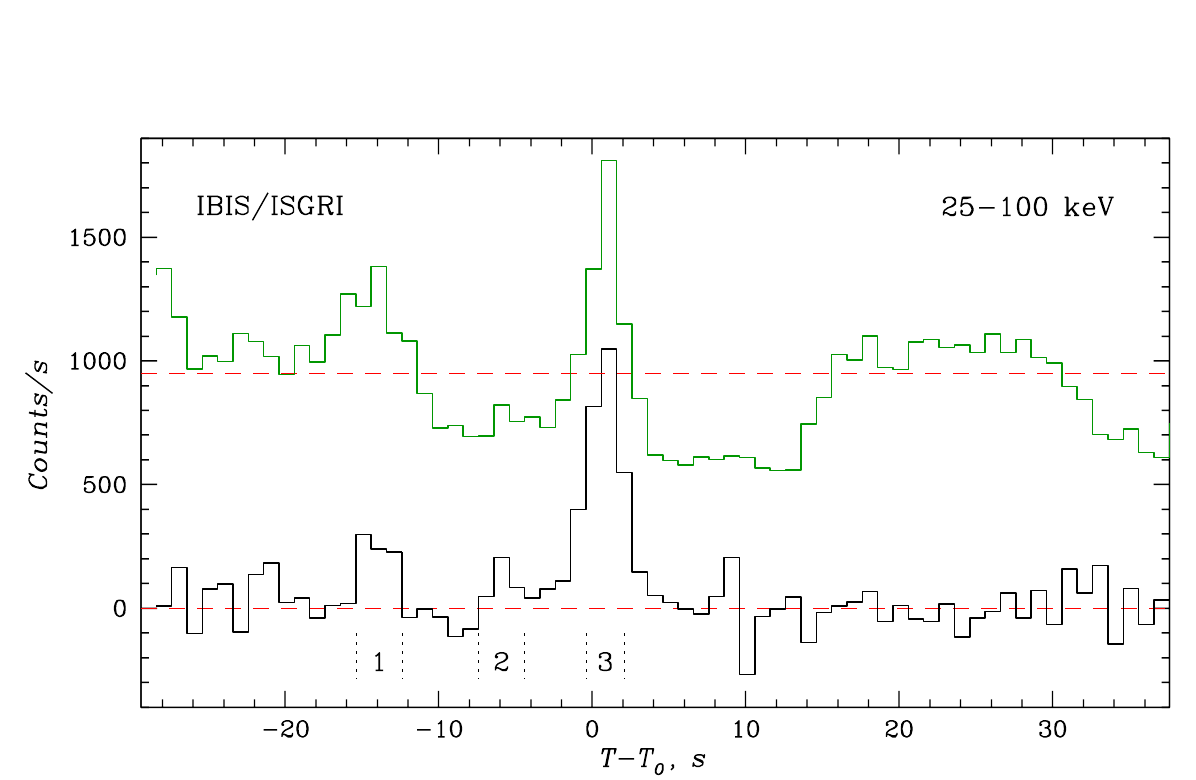}
    \caption{ 20-100 keV count rate history recorded by the IBIS/ISGRI
    detector with  a time resolution of 1.0~s  (green histogram) and the
    corresponding light curve of GRB~210312B reconstructed through the mask
    decoding (black histogram) with the background subtracted and the dead time
    corrected. The vertical dotted lines show time intervals used for image and
    spectral reconstruction for the GRB precursor (1), intermediate (2), and
    primary (3) pulses. \label{fig:lcurve}}
\end{figure}

\begin{figure} \centering
    \includegraphics[width=0.9\hsize]{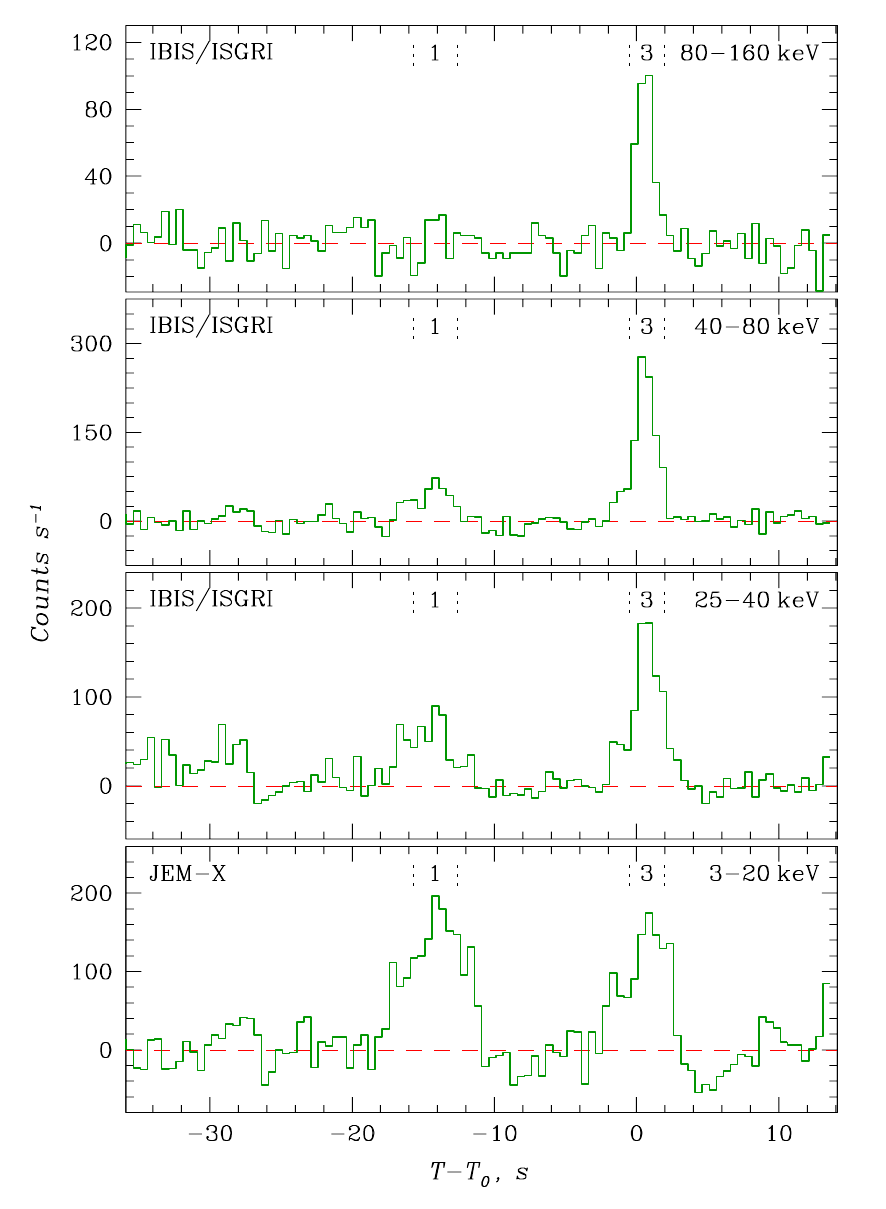}
    \caption{Masked light curves of GRB\,210312B obtained in different energy
    bands with the INTEGRAL IBIS/ISGRI and JEM-X telescopes. The time
    resolution is  0.5\,s. The vertical dashed lines show the time intervals
    used for image and spectral reconstruction for the main pulse and precursor.
    \label{fig:lcmult}}
\end{figure}

\paragraph{Light curve}The light curve of GRB\,210312B obtained by IBIS/ISGRI in the broad
20-100 keV energy range is presented in Fig.~\ref{fig:lcurve}.
The burst occurred when the spacecraft was entering into the Earth's radiation
belts thus the count rate of both instruments is strongly affected by background
variation (green histogram in Fig.\,\ref{fig:lcurve}). The coded mask of
the instrument allows such a variable background to be effectively subtracted
(black histogram). The reconstructed light curve demonstrates the presence of
two additional pulses (indicated in the figure by numbers 1 and 2) preceding the
primary pulse of GRB~210312B (number 3) by $\sim15$ and  $\sim6$~s. Our
reconstruction of the sky image for the 1st pulse confirms its confidence,
$S/N\simeq 6.6\ \sigma$ (see Fig.\,\ref{fig:imgcr}b, left panel). The second pulse
is insufficiently significant, $S/N\simeq 3.1\sigma$, and we did not take it into
account further. Note that the duration of the primary pulse is taken to be only
$2.4$\,s for further analysis although it seems broader in the figure, as this
interval provides the highest S/N for the pulse ($\simeq19.1 \sigma$).
The 1st pulse (precursor) has a duration of $\sim4$~s. Its presence has been
confirmed also by JEM-X in the 3-20 keV range (see Fig.\,\ref{fig:imgcr}, right
panel).
The resulting background-subtracted light curves were obtained by using only
masked (illuminated by the source) pixels of detectors  in three IBIS/ISGRI
energy bands, $25-40$, $40-80$, and $80-160$ keV, and in the \mbox{JEM-X}
$3-20$ keV band (combining the data from the two JEM-X units) are presented in
Fig.~\ref{fig:lcmult}.

\paragraph{Precursor}

After subtracting the background we find another episode of activity in the
IBIS/ISGRI light curves preceding the main pulse (trigger time) by $\sim 17$\,s
(see Fig.~\ref{fig:lcmult}). This emission pulse is softer than the main one: it
is not visible in the hard energy band covering $80-160$ keV, but is actually
brighter than the main episode in the softest energy band covering $3-20$ keV in
the JEM-X data. We calculate hardness ratios between $40-80$ and $3-20$ keV
bands of ISGRI and JEM-X, correspondingly, for both emission episodes, using
their fluxes, expressed in counts. We obtain $HR = 0.37 \pm 0.12$ for the
precursor and $HR = 1.9 \pm 0.4$ for the main episode, confirming the suggestion
of the difference in energy spectra at $4\sigma$ confidence level.

The softness, faintness and relative position of the component suggest it to be
a precursor
\citep[see, e.g.,][and references therein]{min17}. This emission
episode is confirmed by the image reconstruction for both IBIS/ISGRI and JEM-X
instruments, as shown in Fig.~\ref{fig:imgcr}.

\paragraph{Duration}

Using the method described by \citet{kos96}, we estimated the GRB duration
parameters $T_\text{90}$ and $T_\text{50}$ (time intervals with integrated
counts rising from 5\% to 95\% of the total, and from 25\% to 75\%,
correspondingly) in the  $20-160$ keV energy range of the IBIS/ISGRI data:
$T_\text{90} = 18.6\pm 1.5$ s and $T_\text{50} = 2.8 \pm 0.2$ s. The large
difference in the values is connected with the fact that $T_\text{90}$ covers
both episodes of emission, while $T_\text{50}$ covers only the main pulse.
The duration classifies GRB~210312B as a Type II burst (long-duration GRB,
\citealt{Zhang2009ApJ,Kann2011ApJ}).

\subsection{Spectral lag analysis}

Spectral evolution of GRBs is usually measured in terms of a spectral lag -- a
relative shift (lag) of the light curves in different energy bands. The lag is
considered positive if the hard emission is ahead of the soft, and it could be
significant (up to a few seconds) for Type II (long) bursts
\citep[e.g.,][]{Liang2006ApJ}.

To investigate the spectral lag for GRB\,210312B, we used the cross-correlation
method described by \citet{min12,min14} and applied it to the IBIS/ISGRI and
JEM-X data, which cover only the main episode of the emission (time interval (-2,
6)\,s post trigger) and have a S/N sufficient to carry out this type of
analysis.
We constructed light curves with a time resolution of 0.03~s in four
ISGRI energy bands ($20-40$, $40-60$, $60-90$, and $90-160$ keV) and in the
$3-20$ keV energy band of JEM-X. The ISGRI-based energy band $40-60$ keV was
chosen as the reference relative to which the cross-correlation of the other
bands was carried out.

We marginally detect a positive lag for the main emission episode of the burst
over the whole energy range 3-160 keV, including X-rays, which is well described
by a logarithmic function $\textnormal{lag} \propto A\log{E}$ with a spectral
lag index of $A = 0.25 \pm 0.02$. Such a lag also favors the classification of
the burst as a Type II (see, e.g., \citealt{min14}). The null hypothesis of the
absence of the lag is, however, not completely excluded ($\chi^2/\mbox{d.o.f} =
3.3/4$ for zero lag) because of large errors likely related to the faintness of
the burst.


\subsection{X-ray spectra}

\begin{figure} 
    \includegraphics[width=\columnwidth]{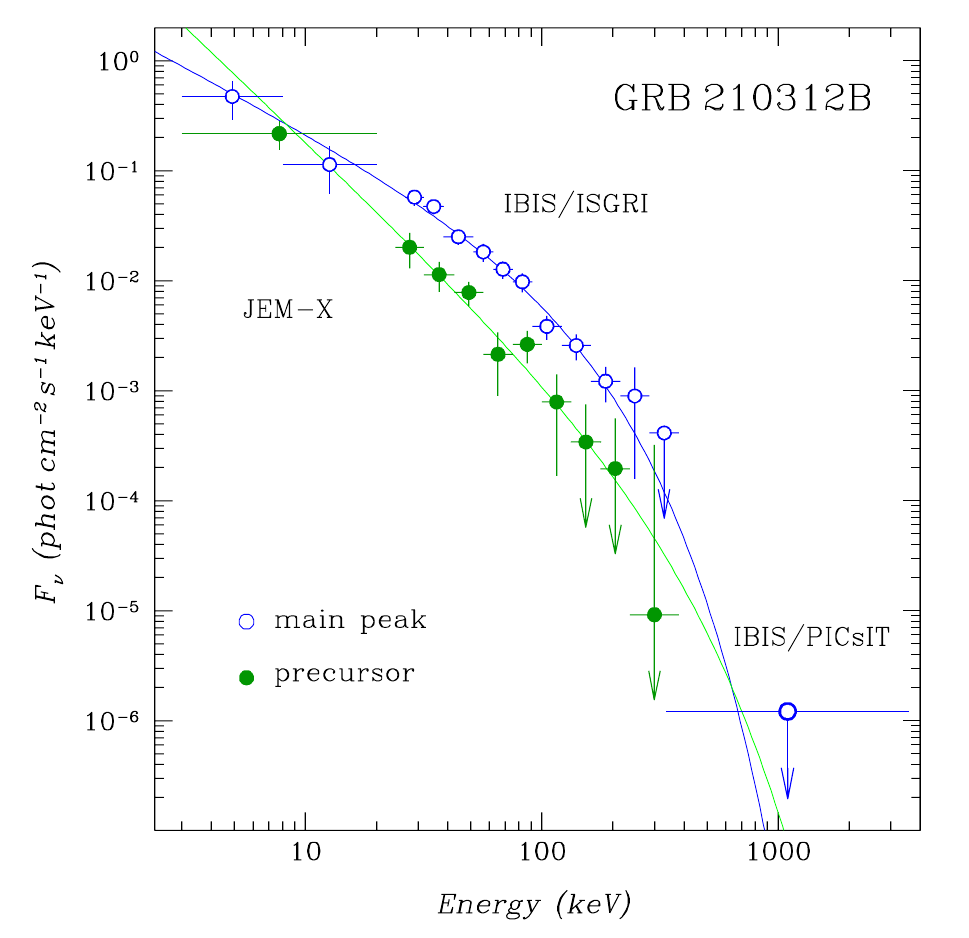}
    \caption{Broadband spectrum measured during the main pulse (blue line
    and open circles) and precursor (green line and filled circles) phases of
    GRB\,210312B as measured with the INTEGRAL IBIS/PICsIT IBIS/ISGRI
    and JEM-X telescopes. The best fits with the CPL model are shown with solid
    lines.\label{fig:spec-PICsIT}}
\end{figure}

The best fit for the main pulse of GRB\,210312B (Fig.~\ref{fig:spec-PICsIT},
open circles) is achieved with the XSPEC Cutoff Power-law (CPL) model. The fit
parameters are: photon index $\alpha=1.2\pm0.2$, cutoff energy $E_{\rm
c}=92\pm31$ keV, $E_{\rm p}=(2-\alpha)E_{\rm c}\simeq74$ keV, and the mean flux
in the 3-400 keV range $F_3\simeq(2.60\pm0.18) \times10^{-7}\ \mbox{erg s}^{-1}
\mbox{cm}^{-2}$. The spectrum of the precursor (Fig.~\ref{fig:spec-PICsIT},
filled circles) is noticeably softer than the spectrum of the main pulse. The
CPL-fit parameters are $\alpha = 2.0\pm 0.3,\ E_{c} \geq 74$ keV (at $3-\sigma$
level) and the mean flux $F_1 = (1.06\pm 0.15)\times 10^{-7}\ \mbox{erg s}^{-1}
\mbox{cm}^{-2}.$ As observed in the optical, the afterglow of GRB\,210312B
reached a maximum $\sim 75$\,s after the GRB trigger time. We therefore
inspected the flux in the ISGRI $25-40$ keV band in the time interval
$T_0+65\,\mbox{s}<T<T_0+85$\,s, and obtained a limit of $F < 6.1\times10^{-10}\
\mbox{erg s}^{-1} \mbox{cm}^{-2}$. \label{sect:xspectra} The JEM-X background
is more sensitive to the radiation environment and was less restricting.

\subsection{$T_\text{90,i}$ -- $EH$ diagram}

To help with the classification problem of GRBs, an advanced method was
proposed by \citet{min20a}. This method uses the $ E_\text{p,i} $ -- $
E_\text{iso} $ correlation features and the bimodality of the rest-frame
duration ($ T_\text{90,i}$) distribution. The $ EH $ parameter
(Eq.~\ref{eq:EH}) characterizes the position of a GRB in the $ E_\text{p,i} -
E_\text{iso} $ diagram, where $ E_\text{iso} $ is an isotropic equivalent of
the total energy emitted in the gamma-ray range, and $ E_\text{p,i} $ is the
position of the maximum in the energy spectrum $ \nu F_{\nu} $ in the source
frame:

\begin{equation}
    EH = \frac{(E_\text{p,i} / 100~\text{keV})}{(E_\text{iso} / 10^{51}~\text{erg})^{~0.4}}.
        \label{eq:EH}
\end{equation}

Using the values of fluence and $ E_\text{p} $ of the CPL spectral model
obtained for the main pulse of GRB\,210312B, and the redshift $z = 1.069$, we
obtain $ E_\text{iso} = (1.70 \pm 0.11)\times10^{51}$ erg and $ E_\text{p,i} =
163 \pm 56$ keV. The corresponding value of the $EH$ parameter is equal to
1.32, while the burst duration in the rest frame is $ T_\text{90,i} = 9.0$\,s.

\begin{figure} 
    \includegraphics[width=\columnwidth]{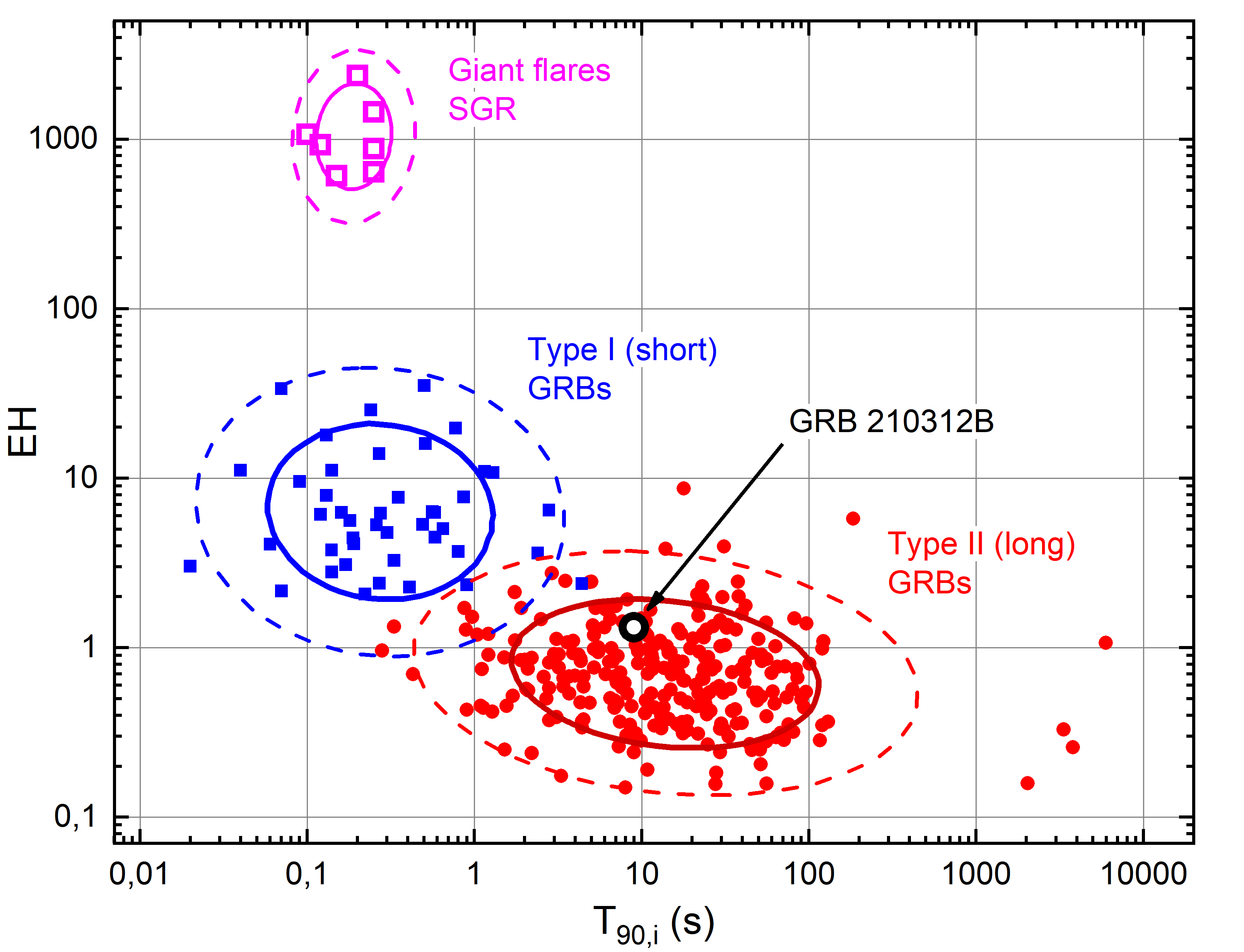}
    \caption{$T_\text{90,i} $ -- $ EH $ diagram for type I GRBs (blue squares),
    type II GRBs (red circles), and SGR giant flares (unfilled magenta  squares)
    with corresponding cluster analysis results. The 1$\sigma $ and 2$\sigma$
    cluster regions are shown with bold solid and thin dashed curves of the
    corresponding colors. GRB\,210312B is shown as an unfilled black circle.
    \label{fig:ehd}}
\end{figure}

\begin{figure} 
    \includegraphics[width=\columnwidth]{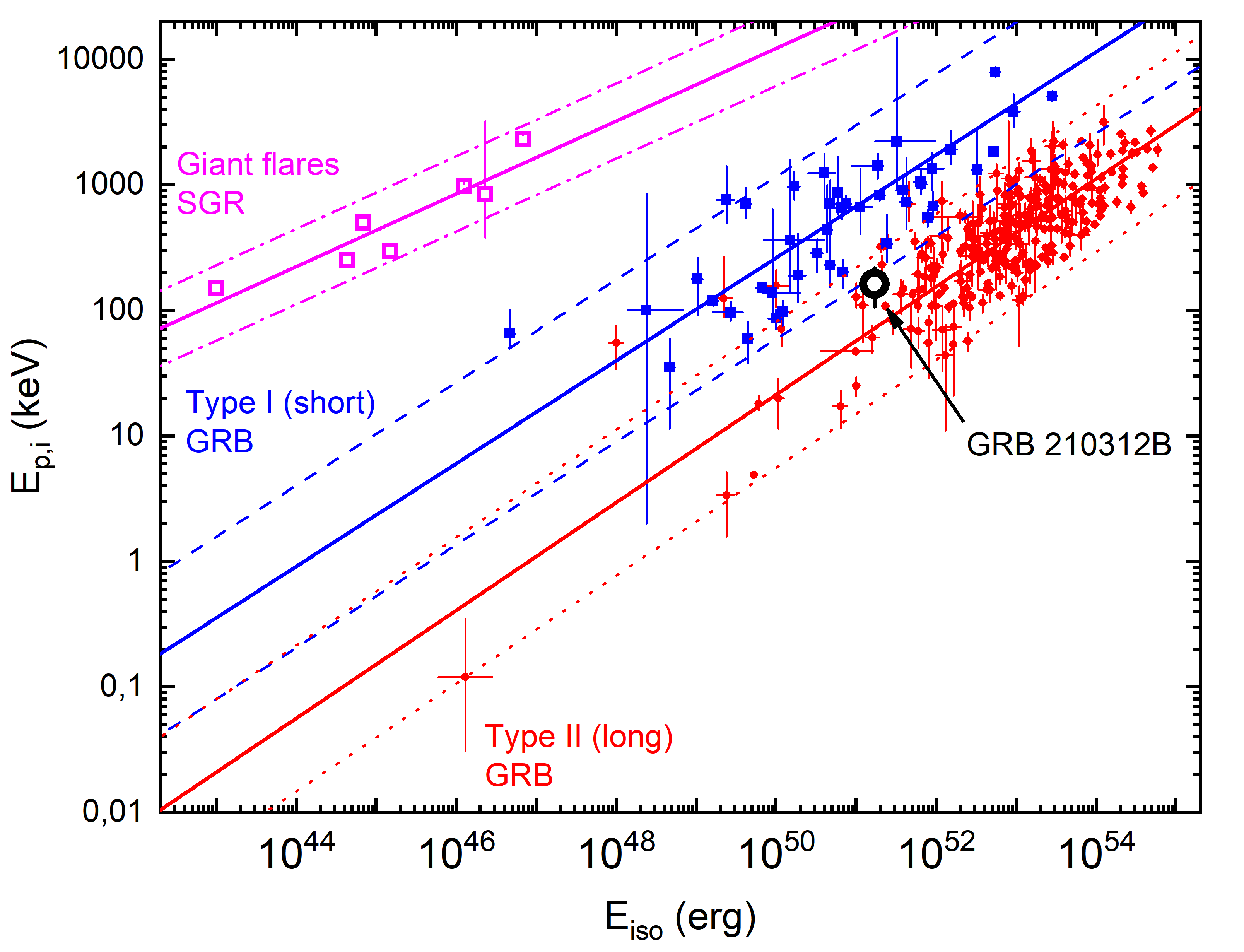}
    \caption{GRB\,210312B in the Amati correlation plane. The location of the
    main pulse is marked. The correlations for long GRBs (Type II; red), short
    GRBs (Type I; blue), and SGR giant flares (purple) are shown with their
    respective $3\sigma$ confidence regions. \label{fig:amati} }
\end{figure}

Type I GRBs, in comparison with Type II GRBs, have a harder spectrum (in terms
of $ E_\text{p,i} $ values) with a lower value of the total isotropic energy $
E_\text{iso} $ and, as a consequence, a larger value of the parameter $ EH$,
and also a shorter duration $ T_\text{90,i}$. As shown in Fig.~\ref{fig:ehd},
GRB\,210312B falls inside the 1$\sigma$ cluster region of Type II bursts,
further supporting our classification. Similarly, its position in the original
Amati $E_{p,i}-E_{iso}$ diagram (Fig.~\ref{fig:amati}) is also consistent with
the Type II origin of the burst.

\subsection{Optical afterglow}

The imaging started 23.8\,s after the GRB. After two initial frames with
low-significance detections, the afterglow is clearly identified rising rapidly
to a bright peak around 75\,s post trigger. The light curve then shows several
distinct phases: an initial decay lasting until $\sim$200\,s, is followed by a
plateau phase for about 40 minutes, a rebrightening around 3000\,s, and finally
by a steady decay. The afterglow remains detectable until $\sim$100\,ks post
trigger, after which it becomes dominated by the host galaxy component.

The photometry requires different approaches at different epochs. In early,
low-resolution images, where the afterglow and both components of the host
system fall within a single point spread function (PSF), we measured their combined brightness and
handled the host contribution through the afterglow modeling. For later,
high-resolution images, we employed PSF-matching image subtraction (hot pants;
\citealt{hotpants}) to isolate the afterglow. Both types of measurements are
indicated accordingly in Appendix\,\ref{tab:phot}.

\subsection{Optical afterglow fitting}

We modeled the optical afterglow using Markov chain Monte Carlo (MCMC) sampling
(\citealt{mcmc1953}) to fit a physically motivated two-component model
consisting of an early optical flare and forward shock emission. The analysis,
performed with 128 walkers over 131072 steps after burn-in, reveals a complex
afterglow evolution characterized by several transitions. Both Akaike and
Bayesian information criteria support our model choice and suggest there are
conservative error estimates in our photometry.

The early optical flare shows a well-defined peak at $T_\mathrm{flare} =
76.0_{-5.1}^{+4.4}$\,s, marked by steep rise ($\alpha_\mathrm{flare,1} =
-4.1_{-2.3}^{+1.6}$) and decay ($\alpha_\mathrm{flare,2} = 4.0_{-1.5}^{+2.1}$)
phases. While the uncertainties in these slopes are substantial due to temporal
overlap with the forward shock component, the timing of the peak itself is
robustly determined. The forward shock evolution begins with a rise to a broad
hydrodynamic peak around $T_\mathrm{FS,12} = 146_{-64}^{+95}$\,s, characterized
by a smooth transition ($N_\mathrm{FS,12} = 0.56_{-0.35}^{+0.30}$) as expected
for the onset of the afterglow.

The subsequent evolution shows a relatively slow decay ($\alpha_\mathrm{FS,2} =
0.87_{-0.26}^{+0.43}$), slower than predicted by the standard fireball model,
suggesting continued energy injection. This phase includes a rebrightening at
$T_\mathrm{FS,23} = 2410_{-500}^{+370}$\,s before finally transitioning to the
late decay at $T_\mathrm{FS,34} = 3840_{-430}^{+380}$\,s. Much like bright and
dim patches in a waterfall emerge from underlying turbulent flow without deeper
physical significance, this     rebrightening likely represents a prominent
manifestation of the stochastic energy injection process rather than a
fundamentally distinct physical transition. The final break at
$T_\mathrm{FS,34}$ marks the cessation of energy injection, after which the
afterglow evolution follows the standard fireball model predictions. The
electron distribution index $p = 2.36_{-0.15}^{+0.18}$ remains well constrained
throughout and implies physically consistent spectral and temporal evolution in
the final decay phase ($\beta = -p/2 = -1.18$, $\alpha_\mathrm{FS,5} = (3p-2)/4
= 1.27$).

This analysis captures the temporal behavior of both the early optical flare and
subsequent forward shock emission, while suggesting that the complex plateau
structure, though well fit by our model, should be interpreted with caution
regarding its physical significance. The host extinction remains consistent with
zero ($A_\mathrm{V,host} = -0.073_{-0.078}^{+0.100}$). The overall goodness of
fit (reduced $\chi^2 = 0.52$) and well-constrained values for key physical
parameters ($p$, $B$, $B_\mathrm{flare}$) support the validity of our
two-component interpretation, even while some transition parameters remain less
certain due to their overlapping effects in the light curve.

\begin{table}
    \centering
    \caption{Parameters of the two-component model (forward shock + early flare) 
    fitted to the GRB\,210312B optical afterglow using MCMC sampling.}
    \begin{tabular}{lccc}
    \hline
    Parameter & Value \\
    \hline
    $\alpha_\mathrm{FS,1}$ & $-1.025_{-1.099}^{+0.704}$ \\
    $\alpha_\mathrm{FS,2}$ & $0.868_{-0.262}^{+0.434}$ \\
    $\alpha_\mathrm{FS,3}$ & $-0.874_{-0.994}^{+0.596}$ \\
    $\alpha_\mathrm{flare,1}$ & $-4.080_{-2.290}^{+1.625}$ \\
    $\alpha_\mathrm{flare,2}$ & $4.025_{-1.501}^{+2.123}$ \\
    $T_\mathrm{FS,12}$ & $145.872_{-64.374}^{+95.314}$ \\
    $T_\mathrm{FS,23}$ & $2407.173_{-503.032}^{+369.532}$ \\
    $T_\mathrm{FS,34}$ & $3836.146_{-425.868}^{+383.336}$ \\
    $T_\mathrm{flare,12}$ & $76.003_{-5.053}^{+4.376}$ \\
    $N_\mathrm{FS,12}$ & $0.561_{-0.353}^{+0.298}$ \\
    $N_\mathrm{FS,23}$ & $0.207_{-0.151}^{+0.193}$ \\
    $N_\mathrm{FS,34}$ & $0.115_{-0.078}^{+0.134}$ \\
    $B$ & $22.066_{-0.337}^{+0.259}$ \\
    $B_\mathrm{flare}$ & $18.227_{-0.315}^{+0.316}$ \\
    $p$ & $2.355_{-0.151}^{+0.175}$ \\
    $A_\mathrm{V,host}$ & $-0.073_{-0.078}^{+0.100}$ \\
    \hline
    \multicolumn{2}{l}{\textit{Derived Parameters:}} \\
    $\alpha_\mathrm{FS,4}$ & $1.266_{-0.113}^{+0.131}$ \\
    $\beta$ & $-1.177_{-0.087}^{+0.076}$ \\
    \hline
    \multicolumn{2}{l}{\textit{Fit Statistics:}} \\
    Number of parameters ($k$) & 16 \\
    Number of data points ($n$) & 38 \\
    Degrees of freedom & 22 \\
    Maximum log-likelihood & -6.97 \\
    Akaike information criterion & 45.94 \\
    Bayesian information criterion & 72.14 \\
    Reduced $\chi^2$ & 0.52 \\
    \hline
    \end{tabular}
    \tablefoot{The model includes temporal evolution with multiple breaks
    ($T_{FS,12-34}$, $T_\mathrm{flare,12}$) and smoothness parameters
    ($N_{FS,12-34}$), spectral evolution governed by electron distribution index
    $p$, and host galaxy extinction $A_\mathrm{V,host}$. $\alpha_{FS,1-4}$ and
    $\alpha_{RS,1-2}$ represent temporal decay indices, while $B$ and
    $B_\mathrm{flare}$ are magnitude zero points at reference times
    ($T_\mathrm{flare,12}$ and 10\,ks, respectively). The quoted uncertainties
    represent the 16th and 84th percentiles of the posterior distributions.
    \label{tab:mcmc_results}}
\end{table}

\begin{figure}
    \centering
    \includegraphics[width=\columnwidth]{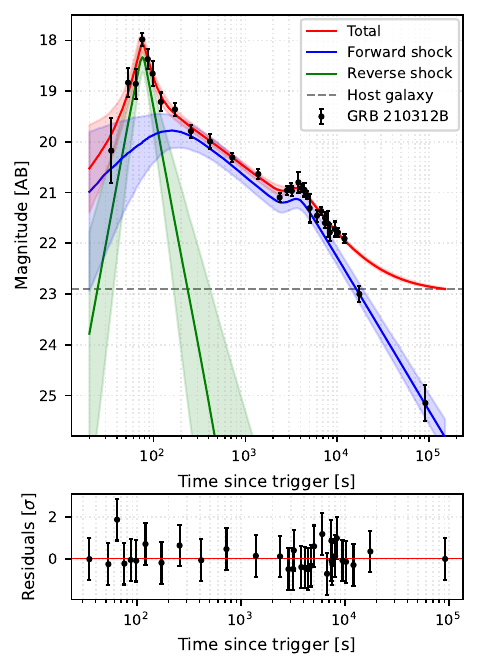}\label{fig:altfit}
    \caption{Optical light curve of GRB\,210312B. The data points show different
    filter measurements converted to the $r$-band equivalent using the spectral
    slope derived from our model fit. The solid lines show the best-fit model
    (red) with 68\% confidence intervals (shaded regions) decomposed into its
    constituent components: forward shock (blue) and an early flare (green)
    emission. The dashed gray line indicates the host galaxy brightness. The
    model simultaneously fits both the temporal evolution and spectral behavior,
    properly accounting for the host galaxy contribution in measurements where it
    was not subtracted through image analysis. The residuals in the bottom panel
    demonstrate the quality of the fit. For model parameters, see
    Table \ref{tab:mcmc_results}. \label{fig:fit}}
\end{figure}

\begin{figure} 
   \centering 
   \includegraphics{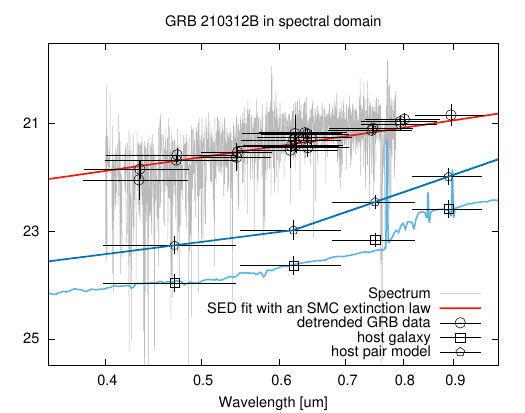}
   \caption{Spectral energy distribution of GRB~210312B and the residuals
   from the temporal and spectral fit model presented in Fig. \ref{fig:fit}.
	The
   points show the data after the temporal evolution is removed, revealing the
   intrinsic spectral shape. The $g r i z$ photometric measurements of both host
   galaxy components are shown (points), along with a representative galaxy
   template spectrum (solid line) for comparison. The dashed line shows our
   simplified model of the host system used for host contribution corrections
   during afterglow fitting. \label{fig:opt_sed}}
\end{figure}

\subsection{Afterglow spectral energy distribution}

The early optical peak reaches $r_\mathrm{AB} = 18.2\,\textnormal{mag}$
($190\,\mu$Jy) at 75\,s post trigger. A contemporaneous search for high-energy
emission in the INTEGRAL data (Sect.\,\ref{sect:xspectra}) yields only
an upper limit in the $20-30$\,keV band of F$<6.1 \times 10^{-10}$\,erg
s$^{-1}$\,cm$^{-2}$ ($25\,\mu$Jy) during 65-85\,s post trigger, implying
$\beta_\mathrm{o-\gamma} < -0.2$.

The MCMC analysis provides well-constrained values for both the electron
distribution index $p = 2.36_{-0.15}^{+0.18}$ and the host extinction
$A_\mathrm{V,host} = -0.073_{-0.078}^{+0.100}$. These yield a spectral index
$\beta = -p/2 = -1.18$ and final temporal decay $\alpha_{FS,4} = (3p-2)/4 =
1.27$, consistent with standard afterglow theory for a forward shock in the slow
cooling regime propagating into a constant density medium. The negligible host
extinction supports this being the intrinsic spectral shape of the afterglow.
    
\subsection{Optical afterglow spectrum}

\begin{figure} 
    \centering
   \includegraphics[width=1.08\hsize]{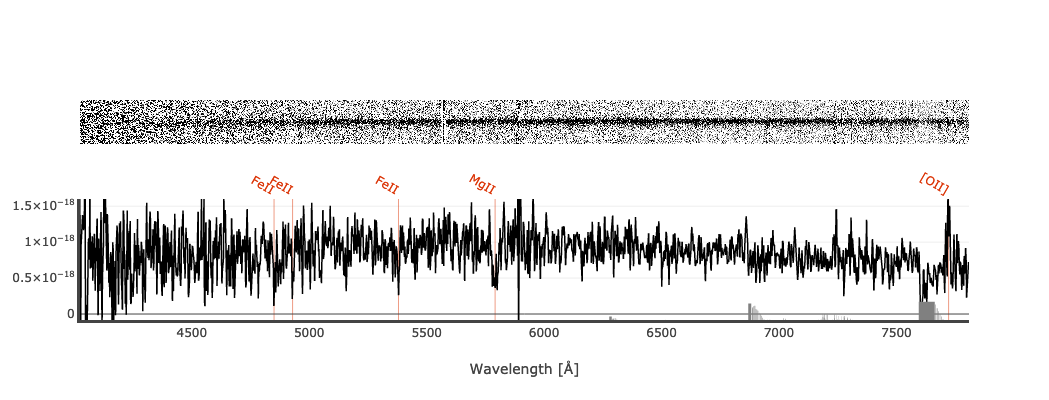}
   \includegraphics[width=\hsize]{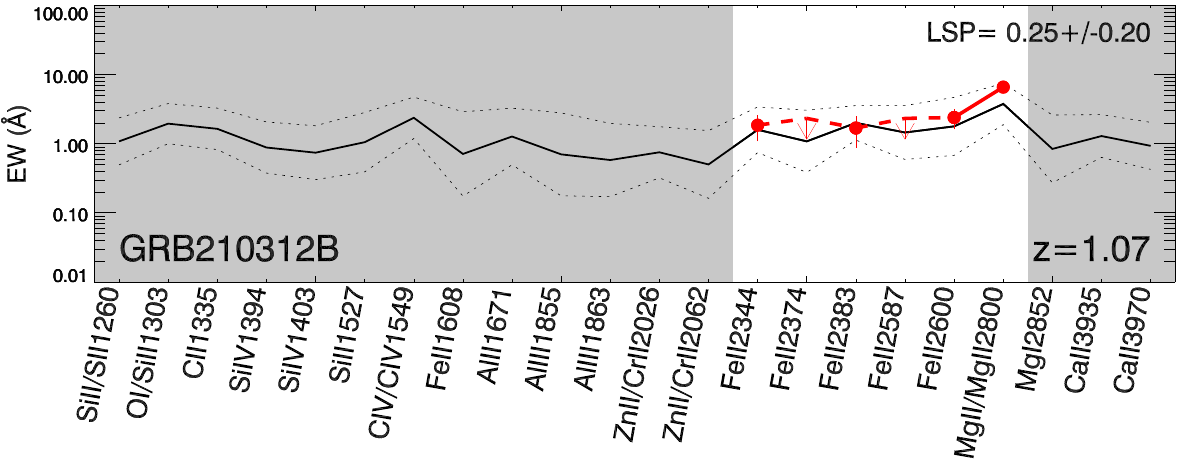}
   \caption{Top: Spectrum obtained with OSIRIS at the 10.4\,m GTC. The 2D and 1D extractions are shown, and the detected
   features are indicated.
   Bottom: Line strength diagram obtained from the
   spectra of GRB\,210312B. \label{fig:spectrum}}
\end{figure}

The spectrum (see Fig. \ref{fig:spectrum}, top) shows a low S/N
continuum in which we can identify several absorption features due to
FeII, MgII, and MgI at a common redshift of $z=1.0690\pm0.0005$. The measurements
of the equivalent widths of the host are presented in Table\,\ref{tab:ews}, with
which we produced a line strength diagram following the prescription of
\cite{deUgarte2012}, which is shown in Fig.\ref{fig:spectrum} (bottom). We
also calculated the line strength parameter of this line of sight, for which we
get a value of $0.25\pm0.20$, which implies that the absorption features
in the line of sight of GRB\,210312B are stronger than 65\% of those in the
sample of \citep{2012A&A...548A..11D}.

We further detect a clear emission feature from the
[OII]$\lambda\lambda3727,3729$ doublet at the same redshift, confirming the
presence of a star-forming host galaxy. The [OII] doublet has a combined flux of
9.1$\pm$0.2$\times$10$^{-18}$ erg\,cm$^{-2}$\,s$^{-1}$, translating to a star
formation rate (SFR) of 0.8\,M$_\odot$\,yr$^{-1}$ \citep{1993RMxAA..27...21K}. This
provides a lower limit since extinction in the host galaxy cannot be reliably
determined from our data. 

We can also estimate the SFR from the rest-frame UV luminosity
of the host galaxy. At $z=1.069$, our $i$-band observations
($\lambda_\mathrm{eff} \approx 7860$\,\AA) correspond to the rest-frame $u$ band.
Using the host galaxy magnitude $i = 23.16\pm0.18$, the solar $u$-band absolute
magnitude from \citet{willmer18}, and the UV--SFR calibration from
\citet{hopkins03}, we derive SFR$_\mathrm{UV} \approx
2.6\,\mathrm{M}_\odot$\,yr$^{-1}$. 


Both estimates suggest modest star formation activity compared to typical rates
of 1-10\,M$_\odot$\,yr$^{-1}$ seen in GRB hosts at similar redshifts. The factor
of 3 difference between UV and [OII] estimates, combined with the relatively
high luminosity and possible interacting nature of the host system, suggests a
complex star formation history that may not be fully captured by either
indicator alone.

\begin{table}
\begin{center}
\caption{Equivalent widths of the absorption features measured in the spectrum.}
\begin{tabular}{cccc} 
Observed $\lambda$ [\AA] & Feature & $z$ & EW [\AA] \\ 
(\AA)   &   &   &   (\AA) \\
\hline  \hline
4850.31 & FeII 2344.21 & 1.0691 & 3.83$\pm$1.57 \\ 
4929.27 & FeII 2382.77 & 1.0687 & 3.48$\pm$1.67 \\
5380.1 & FeII 2600.17 & 1.0691 & 4.95$\pm$1.6 \\ 
5791.49 & MgII 2796.35 & 1.069 & 13.62$\pm$1.88 \\
 & MgII 2803.53 & 1.069 &  \\ \hline
\end{tabular}
\tablefoot{All the wavelengths are in a vacuum.\label{tab:ews}}
\end{center}
\end{table}

\subsection{Host galaxy}
\label{sect:host}

Thirty days after the burst, we reimaged its position with the ASÚ 2\,m
telescope in Ondřejov in the r band with an exposure time of 1.8\,h. The combined
image shows an object whose centroid is not perfectly aligned with the optical
afterglow position and with a brightness of $r = 22.84\pm0.06$,mag.

In order to better study the host galaxy, 52 days after the GRB, we obtained
late imaging with the Large Binocular Telescope (LBT) in $griz$. All four bands
show two objects of comparable brightness (magnitude difference $\simeq0.2$\,mag
in $r$) and color with a relative offset of 1\farcs7 (see
Fig. \ref{fig:opt_image}). GRB\,210312B is located closer to the northern object than the southern one.
With the GRB redshift of $z=1.069$,
observed filter $z$ corresponds to the rest-frame $B$ band and the absolute magnitude
derived from observations in this filter is $M_\mathrm{B} \sim -21.7$, similar
to the Andromeda galaxy. This is notably more luminous than typical GRB hosts at
this redshift \citep{savaglio09, chrimes19}, which tend to be sub-$L^*$ galaxies
with absolute magnitudes around $M_\mathrm{B} \sim -20$. The high luminosity is
also consistent with findings from complete GRB host samples that show most GRB
hosts are typically less luminous at this redshift \citep{perley16}.

According to the DESI Legacy imaging catalog
\citep{legacy}\footnote{https://www.legacysurvey.org/}, which contains the two objects, the northern one is extended and the southern has a stellar PSF. The
brightness of the two objects agrees within error bars with our measurements. Since
the photometry of the system is consistent at four different epochs (Legacy,
CAHA day 4, Ondřejov, LBT), we can place upper limits on any accompanying
supernova using our late-time observations. At the time of the LBT observation
(rest-frame $\sim$25 days), a supernova would need to be brighter than $M_B =
-22.0$ (3$\sigma$ above the host brightness) to be detectable in the $z$ band.
Similarly, from the ASU 2.0m observation (rest-frame $\sim$14 days), we derive a
limit of $M_U = -21.65$ in the observed $r$ band. These limits are well above
the brightest known GRB supernovae, which peak around $M = -20.1$
\citep{Belkin2020AstL...46..783B}, indicating that at this redshift, any
supernova component would be overwhelmed by the host galaxy's brightness.

If indeed the two objects are at the same redshift, they are very likely
interacting  with a distance of only 1.7\arcsec, corresponding to a physical
separation of 11.5\,kpc at $z = 1.069$. The system could be an interacting pair
of galaxies similar to M51 in the local Universe. While most GRB hosts at this
redshift tend to be irregular and show signs of ongoing star formation, the
presence of an interacting system is not unprecedented; several other GRBs
have been found in galaxies that show evidence of interaction
\citep[e.g.,][]{Volnova2014MNRAS.442.2586V} or merging. 

\section{Discussion}

\label{sect:discussion}

\subsection{High-energy domain}

 GRB\,210312B provides an intriguing case study of prompt
emission complexity, detected as a long-duration GRB by INTEGRAL. Its
high-energy light curve revealed two distinct emission episodes separated by
$\sim15$\,s, with the first episode identified as a precursor only after careful
background analysis. Apart from detection by the triggering instrument
IBIS/ISGRI, it joins the select group of GRBs detected by the X-ray telescope
JEM-X \citep{min12, mart13}, providing valuable broad-band coverage of the
prompt emission. The precursor, initially missed in real-time analysis due to
highly variable background \citep{mereghetti21}, was confirmed through detailed
time profile analysis and subsequent image reconstruction. Notably, spectral
analysis reveals the precursor to be softer than the main pulse, a
characteristic seen in some other GRBs with precursor activity.

\subsection{Classification}

The physical classification of GRB\,210312B as a long-duration burst (Type II)
is supported by multiple independent lines of evidence. Beyond the traditional
duration and spectral criteria, we find consistent indicators across temporal
properties (time profile, spectral lag), energetics (position in the $T_{90,i}$
- EH diagram), and environmental characteristics (afterglow behavior, host
galaxy properties, and the small spatial offset from the galaxy center).

Within the context of the INTEGRAL FoV's GRB population
\citep{vianello09}, GRB\,210312B emerges as a relatively modest event. Its
$T_{90}$ of $18.6 \pm 1.5$ seconds places it in the bottom third of INTEGRAL
burst durations, with 68\% of the sample showing longer durations. Similarly,
75\% of INTEGRAL-detected events have higher fluence. Despite these
characteristics, GRB\,210312B displays properties typical of the long-duration
GRB population as a whole.

\subsection{Optical domain}

The optical afterglow of GRB\,210312B exhibits a complex structure that is best
understood as a combination of an early optical flare, forward shock evolution,
and a period of enhanced activity during the plateau phase (see
Fig.\,\ref{fig:fit}). Our detailed MCMC analysis reveals distinct components and
phases in the afterglow evolution.

The early emission is dominated by a sharp peak, characterized by a remarkably
steep rise ($\alpha_\mathrm{flare,1} = -4.1_{-2.3}^{+1.6}$) to a maximum at
$T_\mathrm{flare} = 76.0_{-5.1}^{+4.4}$\,s, followed by a rapid decay
($\alpha_\mathrm{flare,2} = 4.0_{-1.5}^{+2.1}$). The contemporaneous gamma-ray
limit during this peak suggests that the emission peaked in the optical band.
While this and the steep decay phase might be consistent with reverse shock
emission, the physical origin of the early optical flare remains challenging to
determine conclusively. The distinct peak times between the early flare
($\sim$76\,s) and the forward shock component ($\sim$150\,s) challenge standard
reverse shock models, which typically predict coincident hydrodynamic peak times
for both components when followed by steep decay
\citep{kobayashizhang03,zhang03}. Additionally, the observed decay slope
($\alpha_\mathrm{flare,2} = 4.0_{-1.5}^{+2.1}$) is steeper than the theoretical
prediction for reverse shock emission based on our measured electron
distribution index ($\alpha_\mathrm{flare} = (27p+7)/35 \simeq 2.0\pm0.1$).
These discrepancies suggest either a more complex emission scenario or
potentially an internal engine origin for the early flare.

The subsequent plateau phase, lasting until $\sim$2500\,s, shows complex
behavior that cannot be explained by simple relativistic fireball expansion,
suggesting continued energy injection or internal engine activity. This phase
culminates in a rebrightening episode peaking at $T_{FS,34} =
3840_{-430}^{+380}$\,s. While statistically significant, with $\Delta t/t \ge
0.71$ \citep{Swenson2013}, this feature appears to be part of the overall
plateau structure rather than a distinct flaring event. The interpretation of
the second peak as part of the plateau structure is supported by several
observations: the modest amplitude ($\sim$0.25\,mag above the previous minimum),
the smooth transition to the final decay phase, and the lack of significant
spectral evolution during this period.

The afterglow ultimately settles into a well-behaved decay phase characterized
by $\alpha_{FS,4} = 1.27$, consistent with the expectations from standard
afterglow theory for $p = 2.36_{-0.15}^{+0.18}$ (see Sect.\,\ref{sect:closure}).
The absence of significant host extinction ($A_\mathrm{V,host} =
-0.073_{-0.078}^{+0.100}$) and spectral evolution during the late phases
provides additional support for this interpretation of the afterglow behavior.
While we do not observe a jet break in our data, our observations extend only to
$\sim$1 day post-burst, so this non-detection is not strongly constraining as
jet breaks have been observed both before and well after this timescale
\citep[see][]{Li15}.

\subsection{Closure relations}
\label{sect:closure}

The MCMC analysis provides well-constrained values for the temporal and spectral
evolution of the afterglow. The final temporal decay rate $\alpha_\mathrm{FS,4}
= 1.27$ and electron distribution index $p = 2.36_{-0.15}^{+0.18}$ are fully
consistent with the expectations for an adiabatic expansion into a homogeneous
ISM medium in the slow cooling regime (for a complete reference of closure
relations, see \citealt{2013NewAR..57..141G}). The spectral index $\beta = -p/2 =
-1.18$ derived from this electron distribution aligns with standard external
shock model predictions.

The robustness of our interpretation is further supported by independent
analysis. Our detailed analysis is compatible with the preliminary findings
reported by \citet{Kann2024}, who analyzed this burst as part of a larger
sample. Their late-time decay index $\alpha_2 = 1.18 \pm 0.09$ agrees well with
our final temporal decay rate $\alpha_\mathrm{FS,4} = 1.27$, and their spectral
index $\beta = 1.18 \pm 0.077$ is consistent with our value $\beta = -1.18$
derived from the electron distribution index $p = 2.36_{-0.15}^{+0.18}$. The
agreement between these independently derived parameters, with our values
emerging from a physically motivated model constrained by a single $p$ value,
strengthens confidence in the interpretation. Their finding of negative
extinction is also consistent with our results ($A_\mathrm{V,host} =
-0.073_{-0.078}^{+0.100}$). Notably, they identify GRB\,210312B as one of the
faintest afterglows ever followed in such detail, with $R_C \approx 24.7$ mag at
one day post-burst, making our comprehensive early-time coverage particularly
valuable for understanding the prompt-to-afterglow transition in weak bursts.

The negligible host extinction ($A_\mathrm{V,host} = -0.073_{-0.078}^{+0.100}$)
suggests we are observing the intrinsic spectral shape of the afterglow, lending
additional confidence to our derived parameters. This comprehensive consistency
between temporal and spectral properties strongly supports the standard external
shock model as the source of both early and late afterglow emission.

\subsection{Initial gamma factor}

The forward shock peak time $t_{p} = 146_{-64}^{+95}$ s corresponds to the
deceleration time. Since this time is much longer than T$_{90}$ ($\sim$
18.6\,s), it is appropriate to consider a ``thin'' shell regime \citep{sari95}.
The peak time can be used to estimate the initial Lorentz factor of the ejecta
\citep{Meszaros2006,Molinari2007}: 
\begin{equation}
    \Gamma_0 \sim 560(E_{\gamma,iso,52}/\eta_{0.2}n_0t^3_{p,z,1})^{1/8}
,\end{equation}
where $E_{\gamma,iso,52}$ is the isotropic equivalent energy in units of
$10^{52}$ erg s$^{-1}$; $\eta_{0.2}$ is the radiative efficiency in units of
0.2; $n_0$ is local density in units of cm$^{-3}$ and $t_{p,z,1}$ is the peak
time corrected for cosmological time dilation in units of 10\,s. For
GRB\,210312B, with redshift $z = 1.069$ and fluence of $(5.54 \pm 0.35) \times
10^{-7}$ erg cm$^{-2}$ (3-400 keV), we derive the rest-frame isotropic energy
$E_\gamma,iso=(1.7\pm 0.1)\times10^{51}$\,erg. With $t_{p,z,1} =
6.77_{-1.43}^{+1.80}$, we estimate\begin{equation}
    \Gamma_0 = 204_{-13}^{+14}(\eta_{0.2}n_0)^{-1/8}
.\end{equation}
This value is consistent with the established correlation between initial
Lorentz factor and isotropic energy \citep{liang2010}. With this Lorentz factor
and the peak time, the emission radius at the time of maximum is approximately
$R \approx 2c\Gamma^2t/(1+z) \approx 1.7 \times 10^{16}$ cm ($\sim 1100$ AU).

\subsection{Choice of reference time}

The detection of the precursor raises an important question about the choice of
temporal reference ($T_0$) for analyzing the afterglow evolution. While we
primarily used the INTEGRAL trigger time (coincident with the main emission
episode) as our reference, we investigated how shifting $T_0$ to the onset of
the precursor ($\Delta T_0 = -17$\,s) would affect the analysis.

The choice of $T_0$ most significantly impacts the early afterglow temporal
indices. For example, the early flare rise becomes steeper when measured
relative to the precursor onset ($\alpha_\mathrm{flare,1}$ changes from
$-4.1_{-2.3}^{+1.6}$ to approximately $-4.8$). However, the later temporal
indices, including the critical forward shock decay slope $\alpha_{FS,4}$,
remain largely unchanged due to their much larger time differences from $T_0$.

\section{Conclusions}
\label{sect:conclusions}

GRB\,210312B, discovered by INTEGRAL on March 12, 2021, is an excellent
case study of complex GRB evolution from the precursor to the afterglow. The high-energy
emission consisted of two distinct pulses: a precursor and a main episode
separated by approximately 17 seconds, with the precursor showing distinctly
softer emission (hardness ratio $0.37 \pm 0.12$ versus $1.9 \pm 0.4$ for the
main pulse).

The main episode, which lasted $T_{90} = 18.6 \pm 1.5$\,s, is well fit by a cutoff
power law with photon index $\alpha = 1.2 \pm 0.2$ and cutoff energy $E_c = 92
\pm 31$\,keV. A marginally positive spectral lag is detected across the
3-160\,keV range (spectral lag index $A = 0.25 \pm 0.02$). Together with the
burst duration and spectral characteristics, these properties firmly establish
GRB\,210312B as a Type II (long) burst.

Our detailed MCMC analysis of the optical afterglow reveals a complex but
physically coherent evolution.  The early emission shows a prominent optical
flare peaking at $75.8_{-5.1}^{+4.4}$\,s and characterized by steep rise
($\alpha_\mathrm{flare,1} = -4.1_{-2.3}^{+1.6}$) and decay
($\alpha_\mathrm{flare,2} = 4.0_{-1.5}^{+2.1}$) phases. The forward shock
component emerges with a broad hydrodynamic peak around 150\,s, followed by a
plateau phase showing evidence of continued energy injection or engine activity
until about 3800\,s.

The afterglow properties are remarkably consistent with standard external shock
theory. The electron distribution index $p = 2.36_{-0.15}^{+0.18}$ implies
spectral and temporal indices ($\beta = -1.18$, $\alpha_{FS,4} = 1.27$) that
align with expectations for an adiabatic shock in a constant density medium. The
negligible host extinction ($A_\mathrm{V,host} = -0.073_{-0.078}^{+0.100}$)
suggests we are observing the intrinsic afterglow spectrum.

The host appears to be an unusually luminous ($M_B \sim -21.7$), potential
interacting system at $z = 1.069$ with two components separated by 11.5\,kpc.
Spectroscopy reveals both absorption features and [OII] emission, indicating an
actively star-forming environment, though the nature of the southern component
remains uncertain.

 GRB\,210312B stands out for several reasons: the clear detection
of its precursor in multiple energy bands, the well-sampled optical light curve
capturing both the early optical flare and forward shock emission, and the
possibility of decomposing its complex evolution into physically meaningful components
through MCMC analysis. While the lack of X-ray coverage limits our ability to
fully characterize the broadband emission, the available data demonstrate how
careful modeling can extract rich physical information even from partial
spectral coverage.

These results reinforce the importance of rapid response and multiwavelength
observations in modern GRB science, while also highlighting how sophisticated
statistical techniques can help unravel complex afterglow behavior. The
detection of features from the precursor to late afterglow phases makes GRB\,210312B a
valuable addition to the growing sample of well-studied long GRBs.

\begin{acknowledgements}

SG, PM, AP, IC acknowledge support of their analysis of the INTEGRAL data by the
RSCF grant 23-12-00220. DAK acknowledges support from Spanish National Research
Project RTI2018-098104-J-I00 (GRBPhot). AR acknowledges support from the INAF
project Premiale Supporto Arizona \& Italia.

This work is based on observations from multiple facilities. The critical
early-time observations were obtained with the 0.5-m robotic telescope D50 and
the 2-m Perek telescope at the Astronomical Institute of the Czech Academy of
Sciences in Ondřejov, which were instrumental in characterizing the
prompt-to-afterglow transition through rapid-response first detection and
subsequent multi-band monitoring. We used data from the Gran Telescopio Canarias
(GTC), installed at the Spanish Observatorio del Roque de los Muchachos of the
Instituto de Astrofísica de Canarias on the island of La Palma, and from the
2.2m telescope of Centro Astronómico Hispano en Andalucía (CAHA) at Calar Alto
(proposal F21-2.2-013), operated jointly by Junta de Andalucía and Consejo
Superior de Investigaciones Científicas (IAA-CSIC). The LBT observations were
enabled by the international collaboration among institutions in the United
States, Italy, and Germany. The LBT Corporation partners are the University of
Arizona on behalf of the Arizona Board of Regents; Istituto Nazionale di
Astroﬁsica, Italy; LBT Beteiligungsgesellschaft, Germany, representing the
Max-Planck Society, the Leibniz Institute for Astrophysics Potsdam, and
Heidelberg University; and the Ohio State University, representing OSU,
University of Notre Dame, University of Minnesota, and University of Virginia.

\end{acknowledgements}

\begin{table}
    \centering
    \caption{Photometric measurements of the GRB\,210312B host galaxy and its companion 1\farcs7 south of the afterglow position.}
    \begin{tabular}{r c c c l}
    \hline
    exp.\,mid & filter & $T_\mathrm{exp.}$ & mag & telescope \\
    \hline\\
\multicolumn{5}{c}{The nothern underlying object (host galaxy)}\\
\hline
52.2976 & $     g       $ & 10$\times$120\,s    &       23.96$\pm$0.17  &       LBT\,8.4m \\
52.2976 & $     r       $ & 10$\times$120\,s    &       23.64$\pm$0.13    &      LBT\,8.4m \\
52.3216 & $     i       $ & 10$\times$120\,s    &       23.16$\pm$0.18  &       LBT\,8.4m \\
52.3216 & $     z       $ & 10$\times$120\,s    &       22.58$\pm$0.16  &       LBT\,8.4m \\
\hline \\
\multicolumn{5}{c}{The southern underlying object (host companion)}\\
\hline
52.2976 & $     g       $ & 10$\times$120\,s    &       24.07$\pm$0.18  &       LBT\,8.4m \\
52.2976 & $     r       $ & 10$\times$120\,s    &       23.83$\pm$0.13  &       LBT\,8.4m \\
52.3216 & $     i       $ & 10$\times$120\,s    &       23.26$\pm$0.20  &       LBT\,8.4m \\
52.3216 & $     z       $ & 10$\times$120\,s    &       22.91$\pm$0.21  &       LBT\,8.4m \\
\hline
\\
    \end{tabular}
    \tablefoot{Magnitudes are not corrected for Galactic extinction. As above, the values are given in their native systems: Pan-STARRS $griz$ magnitudes are AB, $BVR_C NI_C$ are in Vega.}
    \label{tab:phot1}
\end{table}

%
%
    
\bibliographystyle{aa}
\bibliography{aa53636-24}

\clearpage  
\onecolumn  

\appendix
    \section{Photometric measurements of the GRB\,210312B optical afterglow}
    \centering
    \begin{tabular}{r c c c l}
    \\
    \hline
    exp.\,mid & filter & $T_\mathrm{exp.}$ & mag & telescope \\
(seconds) & & & & \\
    \hline\\
\multicolumn{5}{c}{Afterglow + both underlying objects measured as a single detection}\\
\hline
34.67 & $cr$ & 2$\times$10\,s & {$>19.5$} & ASU\,0.5m \\ 
52.84 & $cr$ & 10\,s & 18.83$\pm$0.28 & ASU\,0.5m \\ 
64.07 & $cr$ & 10\,s & 18.86$\pm$0.29 & ASU\,0.5m \\ 
75.31 & $cr$ & 10\,s & 17.98$\pm$0.13 & ASU\,0.5m \\ 
86.54 & $cr$ & 10\,s & 18.37$\pm$0.19 & ASU\,0.5m \\ 
97.86 & $cr$ & 10\,s & 18.66$\pm$0.25 & ASU\,0.5m \\ 
120.1 & $cr$ & 3$\times$10\,s & 19.21$\pm$0.19 & ASU\,0.5m \\ 
171.3 & $cr$ & 2$\times$10\,s & 19.36$\pm$0.13 & ASU\,0.5m \\ 
255.5 & $cr$ & 9$\times$10\,s & 19.79$\pm$0.13 & ASU\,0.5m \\ 
412.8 & $cr$ & 10$\times$20\,s & 19.99$\pm$0.14 & ASU\,0.5m \\ 
721.9 & $cr$ & 19$\times$20\,s & 20.31$\pm$0.09 & ASU\,0.5m \\ 
1384 & $cr$ & 43$\times$20\,s & 20.63$\pm$0.09 & ASU\,0.5m \\ 
3208 & $cr$ & 129$\times$20\,s & 20.97$\pm$0.10 & ASU\,0.5m \\ 
\hline
5370 & $B$ & 600\,s & 21.94$\pm$0.18 & CAHA\,2.2m \\ 
7508 & $B$ & 5$\times$90\,s & 22.32$\pm$0.37 & OSN\,1.5m \\ 
\hline
3160 & $g$ & 300\,s & 21.18$\pm$0.08 & ASU\,2.0m \\ 
4405 & $g$ & 300\,s & 21.25$\pm$0.13 & ASU\,2.0m \\ 
\hline
6028 & $V$ & 600\,s & 21.58$\pm$0.12 & CAHA\,2.2m \\ 
8021 & $V$ & 1283\,s & 21.74$\pm$0.26 & OSN\,1.5m \\ 
\hline
2235 & $r$ & 120\,s & 21.00$\pm$0.19 & ASU\,2.0m \\ 
2490 & $r$ & 120\,s & 21.22$\pm$0.27 & ASU\,2.0m \\ 
2616 & $r$ & 120\,s & 20.95$\pm$0.18 & ASU\,2.0m \\ 
4713 & $r$ & 300\,s & 21.03$\pm$0.35 & ASU\,2.0m \\ 
5024 & $r$ & 3$\times$300\,s & 21.28$\pm$0.28 & ASU\,2.0m \\ 
7524 & $r$ & 60\,s & 21.50$\pm$0.09 & GTC 10.4m \\ 
2590280 & $r$ & 12$\times$600\,s & 22.98$\pm$0.10 & ASU\,2.0m \\ 
\hline
6699 & $R_C$ & 600\,s & 21.14$\pm$0.07 & CAHA\,2.2m \\ 
8317 & $R_C$ & 5$\times$90\,s & 21.48$\pm$0.17 & OSN\,1.5m \\ 
9426 & $R_C$ & 18$\times$180\,s & 21.48$\pm$0.14 & OSN\,0.9m \\ 
10200 & $R_C$ & 10$\times$180\,s & 21.50$\pm$0.09 & OSN\,1.5m \\ 
12030 & $R_C$ & 10$\times$180\,s & 21.64$\pm$0.09 & OSN\,1.5m \\ 
91500 & $R_C$ &  $10\times500$\,s & 22.89$\pm$0.11 & OSN\,1.5m $^a$ \\ 
209200 & $R_C$ & 32$\times$120\,s &  {$>20.5$} & AZT33\,1.5m  \\ 
340400 & $R_C$ & 6$\times$600\,s & 23.27$\pm$0.20 & CAHA\,2.2m\\ 

\hline
2852 & $i$ & 300\,s & 20.64$\pm$0.10 & ASU\,2.0m \\ 
4097 & $i$ & 300\,s & 20.58$\pm$0.08 & ASU\,2.0m \\ 
\hline
4707 & $I_C$ & 600\,s & 20.26$\pm$0.08 & CAHA\,2.2m \\ 
7376 & $I_C$ & 5$\times$90\,s & 20.78$\pm$0.11 & OSN\,1.5m \\ 
7392 & $I_C$ & 600\,s & 20.68$\pm$0.11 & CAHA\,2.2m \\ 
\hline
3787 & $z$ & 300\,s & 20.19$\pm$0.20 & ASU\,2.0m \\ 
\hline\\
\multicolumn{5}{c}{The afterglow only (image subtraction)}\\
\hline
17480 & $r$ & 240\,s & 23.00$\pm$0.15 & GTC 10.4m  \\ 
91500 & $R_C$ & $10\times500$\,s & 24.93$\pm$0.35 & OSN\,1.5m  \\ 
\hline
\\
\end{tabular}
\tablefoot{Magnitudes are not corrected for Galactic extinction. They are given
in their native systems: Pan-STARRS $griz$ magnitudes are AB, $BVR_C I_C$ are
in Vega. $^a$ the host galaxy contribution was removed by image subtraction and
photometry of the pure afterglow is shown at the lowest part of the table.
\label{tab:phot} }
    

\end{document}